\documentclass[twocolumn,journal]{IEEEtran}

\usepackage{amsmath}
\usepackage{amssymb}
\usepackage{amsfonts}
\usepackage{bm}
\usepackage{dsfont}
\usepackage{bbm}
\usepackage{mathtools}
\usepackage{mathdots}
\usepackage{nccmath}
\usepackage{yhmath}

\usepackage{graphicx}
\usepackage{xcolor}
\usepackage{graphics}
\usepackage{pgfplots} 
\usepackage{psfrag}

\usepackage{array}
\usepackage{bigstrut}
\usepackage{multirow}
\usepackage{tabularx}
\usepackage{booktabs}
\usepackage{colortbl}
\usepackage{diagbox}
\usepackage{multicol}
\usepackage{caption}
\usepackage{subcaption}
\usepackage{float}
\usepackage{balance}
\usepackage{afterpage}

\usepackage{cite}
\usepackage{hyperref}
\usepackage{siunitx}
\usepackage{enumitem}
\usepackage{verbatim}
\usepackage{stmaryrd}

\usepackage{alphabeta}
\usepackage{algpseudocode}
\usepackage{algorithm}
\usepackage{stackengine}
\usepackage{gensymb}

\newtheorem{lemma}{Lemma}

\usepackage{tikz}
\usepackage{pgfplots}
\pgfplotsset{compat=1.15}
\usepackage{mathrsfs}
\usetikzlibrary{arrows}
\usetikzlibrary{decorations.pathreplacing}
\usetikzlibrary{fadings}

\usepackage{tikz}
\usetikzlibrary{quantikz}

\begin{document}
	\IEEEoverridecommandlockouts
	
 \title{Quantum Approximate Optimization Algorithm for MIMO with Quantized $b$-bit Beamforming}
\author{Nikos~A~Mitsiou,~\IEEEmembership{Graduate Student Member,~IEEE}, \\ Ioannis~Krikidis,~\IEEEmembership{Fellow,~IEEE} and George~K.~Karagiannidis,~\IEEEmembership{Fellow,~IEEE}
\thanks{N.~A.~Mitsiou and G.~K.~Karagiannidis are with the Department of Electrical and Computer Engineering, Aristotle University of Thessaloniki, 54124 Thessaloniki, Greece (e-mails: nmitsiou@auth.gr, geokarag@auth.gr).}
\thanks{I. Krikidis is with the Department of Electrical and Computer Engineering, University of Cyprus, 1678 Nicosia, Cyprus (e-mail: krikidis@ucy.ac.cy).}\vspace{-0cm}
\thanks{This project has been supported by the European Research Council (ERC) under the European Union’s Horizon Europe research and innovation programme, grant agreement No. 101241675 (ERC PoC QUARTO).}
}
\maketitle

\begin{abstract}
Multiple-input multiple-output (MIMO) is critical for 6G communication, offering improved spectral efficiency and reliability. However, conventional fully digital designs face significant challenges due to high hardware complexity and power consumption. Low-bit MIMO architectures, such as those employing \(b\)-bit quantized phase shifters, provide a cost-effective alternative but introduce NP-hard combinatorial problems in the pre- and post-coding design. This paper explores the use of the Quantum Approximate Optimization Algorithm (QAOA) and alternating optimization to address the problem of $b$-bit quantized phase shifters both at the transmitter and the receiver. We demonstrate that the structure of this quantized beamforming problem aligns naturally with hybrid-classical methods like QAOA, as the phase shifts used in beamforming can be directly mapped to rotation gates in a quantum circuit. Notably, this paper is the first to show that theoretical connection. \textcolor{black}{Then, the Hamiltonian derivation analysis for the \(b\)-bit case is presented, which could have applications in different fields, such as integrated sensing and communication, and emerging quantum algorithms such as quantum machine learning.} In addition, a warm-start QAOA approach is studied which improves computational efficiency. Numerical results highlight the effectiveness of the proposed methods in achieving an improved quantized beamforming gain over their classical optimization benchmarks from the literature.
\end{abstract}
\vspace{-0cm}

\section{Introduction}
Multiple-input multiple-output (MIMO) systems are a key technology for future 6G wireless communication, utilizing spatial degrees of freedom to address the increasing demands for capacity, throughput, and reliability. However, as the number of antennas grows, conventional fully digital MIMO systems encounter major challenges related to implementation complexity, cost, and power consumption \cite{tut}. These challenges stem from the hardware requirements of radio frequency (RF) chains, digital-to-analog converters (DACs), analog-to-digital converters (ADCs), and baseband signal processing. To overcome these limitations, hybrid analog/digital MIMO architectures have been introduced \cite{wang}, which distribute the processing between the analog and the digital domains, achieving a balance between performance and implementation efficiency. Other strategies include low-resolution phase shifters, which are suitable for high-frequency bands, as well as the use of low-resolution DACs or ADCs \cite{masouros, silva}.

Such a novel MIMO architecture, employing 1-bit signal processing resolution at both the transmitter and receiver, was proposed in \cite{krikidis1,krikidis2} to significantly reduce hardware complexity and power consumption. Unlike conventional 1-bit DAC/ADC architectures of \cite{masouros, wang, silva}, this approach assumes low-resolution pre- and post-coding at both ends of the communication link. This design is particularly promising for low-power and low-computation devices, especially within the Internet of Things (IoT). However, the binary nature of the signal processing introduces a complex NP-hard combinatorial optimization problem in determining the pre- and post-coding vectors.  \textcolor{black}{Specifically, the complex-valued binary resolution of this MIMO architecture leads to an NP optimization problem, whose optimal solution currently admits no known polynomial-time algorithm.}

Quantum computing offers a promising alternative for addressing NP-hard combinatorial optimization problems, with various applications in wireless networks as well \cite{hanzo}. The Quantum Approximate Optimization Algorithm (QAOA) has emerged as a leading candidate for solving such problems on gate-model quantum computers \cite{abbas2023quantum}. Designed to tackle hard optimization tasks like the MAX-CUT problem \cite{guerreschi2019qaoa}, QAOA leverages the power of hybrid quantum-classical computation.  \textcolor{black}{The optimization function is first written as a cost Hamiltonian. Then, a tunable quantum circuit is built that repeatedly evolves under this cost Hamiltonian and a mixer Hamiltonian,} and which is iteratively updated by a classical optimizer, providing flexibility to adapt the algorithm to the constraints of noisy intermediate-scale quantum (NISQ) devices \cite{symons2023practitioner}. 

\subsection{Literature Review}
\textcolor{black}{Recent studies have highlighted the potential of hybrid beamforming for ultra-massive MIMO and sensing systems utilizing THz bands \cite{10891254}. Additionally, dynamic metasurface antennas (DMAs) have emerged as a compact and energy-efficient solution for optimizing beamforming in downlink communication systems \cite{10786283}. Toward energy-efficient and cost-effective beamforming, low-resolution MIMO systems with quantized phase shifters and low-complexity ADCs/DACs have also gained significant attention.} In \cite{ding,liu} highlighted the spectral and energy efficiency of hybrid precoding schemes, addressing the impact of additive quantization noise in mmWave scenarios. In addition,  \cite{dutta} explored digital beamforming with low-resolution ADCs, emphasizing architectural tradeoffs for mmWave systems. Moreover, \cite{ribeiro, Mahmood} advanced the understanding of analog beamforming and hybrid precoding, focusing on energy-efficient designs using low-resolution DACs and 2D antenna arrays. Further, \cite{kim} addressed scalability in cell-free mmWave MIMO systems by integrating low-capacity fronthaul links and analog beamforming, while \cite{choi} demonstrated the viability of coordinated beamforming with quantized hardware. In \cite{chang}, a hybrid beamforming was introduced for terahertz massive MIMO systems, leveraging low-resolution phase shifters and time-delay elements to optimize performance in high-frequency bands.
Furthermore, \cite{krikidis1,krikidis2} proposed a 1-bit signal processing resolution at both the transmitter and the receiver as a means to substantially reduce hardware complexity and power consumption. Unlike traditional 1-bit DAC/ADC designs, this method utilizes low-resolution pre-coding and post-coding at both ends of the communication link. \textcolor{black}{Then, the 1-bit beamforming optimization was formulated, and tackled using quantum annealing (QA), which achieved performance equivalent to the performance of the exhaustive search method.}

In fact, the integration of quantum computing techniques, such as QAOA and Grover's algorithm, into MIMO systems has been actively explored to enhance computational efficiency and tackle optimization challenges. In \cite{Habibie}   Grover's quantum search was utilized for active user detection in IoT networks, highlighting its application in MIMO-OFDM systems. Also, \cite{Ishikawa} showed that Grover's algorithm offers quadratic speedup for index modulation, a key feature for massive MIMO, while \cite{Sano} integrated QAOA for optimizing wireless channel assignments in multi-user systems. In addition, \cite{Cui} explored QAOA for maximum likelihood detection (MLD) in MIMO, demonstrating its capability to achieve improved accuracy and efficiency,  while in \cite{Liu2} an improved QAOA-based approach for massive MIMO MLD was presented.  Additionally, \cite{Matsumine} reviewed the potential of quantum algorithms like Grover and QAOA for physical layer security in massive MIMO, emphasizing their scalability. In \cite{Narottama}, combined QAOA with Grover's algorithm for efficient transmit precoding in MIMO systems. Moreover, \cite{Neumann} highlighted quantum computing applications in MIMO radar systems, utilizing Grover's algorithm to optimize performance. Lastly, \cite{Gülbahar} utilized Grover-based quadratic speedup for MIMO MLD, addressing scalability challenges without requiring quantum hardware with large qubit numbers.

\subsection{Motivation \& Contribution}
Despite the potential of low-bit resolution signal processing for balancing the tradeoff between performance, hardware complexity and power consumption, one challenge which stems from reducing the quantization resolution is that it leads to NP-hard optimization problems. \textcolor{black}{NP-hard optimization problems admit no known polynomial-time solution; under the widely believed conjecture $\text{P}\neq\text{NP}$, any algorithm that always finds an optimal solution must run in super-polynomial (and typically exponential) time in the worst case, although it remains unproven whether exponential time is strictly necessary.} \textcolor{black}{QAOA offers a promising quantum-classical heuristic for finding high-quality approximate solutions.} However, for the wireless and signal processing community, when formulating combinatorial optimization problems to assess the performance of QAOA, the focus has mainly been on problems that naturally admit quadratic objectives, using QAOA as a black-box optimizer. This focus is largely due to the structure of the off-the-shelf quantum optimization frameworks, which often rely on pairwise interactions. However, QAOA does not have this inherent limitation because gate-model quantum machines are capable of implementing higher-order terms (HOTs). This capability effectively expands the problem space to include objectives involving products of more than two binary variables, enabling a broader class of optimization challenges to be addressed. \textcolor{black}{In fact, the incorporation of HOTs is relevant for many emerging technologies, such as integrated sensing and communication and low-resolution signal processing.} 

To demonstrate this capability of QAOA, and its relevance to wireless communications problems, this work extends the analysis presented in \cite{krikidis1,krikidis2}. \textcolor{black}{The work of [5],[6] deals only with the special case $b = 2$ using QA. By contrast, our paper develops a general analysis valid for any $b$ and shows how QAOA can serve as a promising solver. Moreover, the proposed framework can naturally extend to other challenging wireless network applications, such as RIS phase shift optimization, whereas it is unclear whether the solution of [5],[6] could be adapted. Specifically, RIS phase-shift optimization uses continuous phase values that can directly be mapped to rotation gates in a quantum circuit, while the method in [5],[6] is known to tackle only combinatorial problems.} \textcolor{black}{We note that the interest in analyzing the $b$-bit resolution arises since values such as $b = 2, 3$ can offer a significant tradeoff between system performance and implementation complexity.} 
Thus, we leverage the QAOA in combination with an alternating optimization (AO) approach to separately tackle the pre-coding and post-coding optimization problems. We demonstrate that the structure of this quantized beamforming problem naturally aligns with QAOA. Specifically, the phase shifts employed in beamforming can be directly mapped to rotation gates within a quantum circuit, establishing a novel theoretical connection between quantum circuits and phase shifts in pre-coding and post-coding. This paper is the first to explicitly describe this connection, providing a detailed analytical procedure to derive the cost Hamiltonian for the problem. This Hamiltonian is then mapped to the quantum operators required for the QAOA process. Additionally, we utilize a warm-start strategy that improves the accuracy of the solution by initializing the algorithm with a different mixer Hamiltonian. Interestingly, we show that using QAOA as a generic black-box optimizer for the case of $b = 2$ produces suboptimal solutions compared to our tailored approach, further reinforcing the motivation to customize QAOA to the specific characteristics of the problem instance. Numerical results confirm the effectiveness of the proposed QAOA-based solution, highlighting its potential for tackling low-resolution beamforming optimization problems. The contributions of the paper are summarized below:
\begin{itemize}
    \item This work extends prior analyses by formulating and solving the NP-hard beamforming optimization problem for \(b\)-bit quantized pre-coding and post-coding vectors, addressing the tradeoffs between MIMO performance and hardware complexity.

    \item A theoretical connection between quantum circuits and beamforming phase shifts is introduced. Specifically, the phase shifts can be directly mapped to rotation operator gates around the $z$-axis. \textcolor{black}{Then, a detailed derivation of the Hamiltonian corresponding to the $b$-bit beamforming problem is provided. We note that this analysis is relevant to most emerging quantum optimization methods, such as quantum machine learning.} 

    \item Based on this, a hybrid classical-quantum method based on QAOA and AO is proposed, which tackles the quantized beamforming problem. A warm-start strategy is also utilized to enhance solution accuracy and efficiency.

    \item Simulations showcase the efficiency of the proposed approach over conventional methods like quantized singular value decomposition, and its optimal performance as compared to the solution derived from exhaustively searching the whole solution space.
\end{itemize}

\section*{Structure}
The structure of the paper is as follows. Section II introduces the necessary preliminaries on QAOA. Section III describes the investigated \(b\)-bit beamforming system model. Section IV focuses on the QAOA-based beamforming optimization approach for the simple case of \(b = 2\), while Section V extends this framework to the general case of \(b\)-bit quantization. Section VI provides a complexity analysis of the proposed algorithms. Section VII presents numerical simulations to evaluate the performance of the methods, and Section VIII concludes the paper with a summary of the findings and potential future research avenues.
\section*{Basic notation}
The ket notation \(|\psi\rangle\) represents a column vector and its corresponding bra notation \(\langle\psi|\) is the Hermitian conjugate of the ket. The symbol $\mathfrak{i}$ denotes the imaginary unit, while the symbol $i$ is an index. To express quantum states in the computational basis, we define the standard qubit states \( |0\rangle = \begin{pmatrix} 1 \\ 0 \end{pmatrix}\) and \( |1\rangle = \begin{pmatrix} 0 \\ 1 \end{pmatrix}\).
In this notation, \( |0\rangle \) represents the state where the qubit is in the ``zero" state, while \( |1\rangle \) represents the ``one" state.
\textcolor{black}{The Pauli-Z operator \( Z = \begin{pmatrix} 1 & 0 \\ 0 & -1 \end{pmatrix}\) is a fundamental operator in quantum mechanics that measures the \( z \)-axis component of the qubit's spin. Its action on a qubit basis \( |\mathbf{x}\rangle \), where \( x \in \{0, 1\} \), is given by
\begin{equation}
    Z | \mathbf{x} \rangle = (-1)^{x} |\mathbf{x}\rangle,
\end{equation}
which implies that \( Z|0\rangle = |0\rangle \) and \( Z|1\rangle = -|1\rangle \). This property of the Pauli-Z operator is crucial in encoding binary variables in quantum states, as it assigns a phase of \( +1 \) to \( |0\rangle \) and \( -1 \) to \( |1\rangle \), which can be used in optimization and eigenvalue-based calculations. We note that the Pauli-Z operator when acting on the \( k \)-th qubit will be denoted as $Z_k$. The same will hold for any other Pauli operator too.}  
 Also, the Pauli $X$ operator is defined as \(X= \begin{pmatrix} 0 & 1 \\ 1 & 0 \end{pmatrix}\). Bold characters denote vectors, while upper case letters denote either matrices, or operators, based on the context. The symbol notation $^{\otimes n}$ denotes the tensor product of $n$ identical states or operators.
\section{Preliminaries on QAOA}
\subsection{The QAOA method}
The QAOA is a hybrid quantum-classical algorithm designed for combinatorial optimization and quadratic unconstrained binary optimization  (QUBO) problems with a direct connection to the quantum adiabatic theorem \cite{farhi2014quantum}. \textcolor{black}{ QAOA is particularly suited for implementation on near-term quantum devices, often referred to as NISQ devices, due to its low circuit depth and robustness against certain types of noise \cite{zhou2020quantum, farhi2014quantum}.} The connection between QAOA and QUBO problems is direct. The QUBO problem is formulated by considering a binary vector of length \( n \), \( \mathbf{x} \in \{0,1\}^n \), and a square matrix \( Q \in \mathbb{R}^{n \times n} \). The elements \( Q_{ij} \) are weights for each pair of indices \( i, j \) in \( \mathbf{x} \). Consider the function \( f_Q : \mathbb{B}^n \rightarrow \mathbb{R} \), which is given below
\begin{equation} \label{eq:qubo}
f_Q(\mathbf{x}) = \langle \mathbf{x} | Q | \mathbf{x} \rangle + \langle \mathbf{x} | \mathbf{c} \rangle = \sum_{i=1}^n \sum_{j=1}^n Q_{ij} x_i x_j + \sum_{i=1}^n c_i x_i,
\end{equation}
where \( x_i \) is the \( i \)-th element of \( | \mathbf{x} \rangle \). Please note that the notation $\langle \mathbf{x} | Q | \mathbf{x} \rangle$ is equal to the notation $\mathbf{x}^{\dagger}Q\mathbf{x}$, where $\mathbf{x}^{\dagger}$ is the conjugate transpose of vector $\mathbf{x}$. 
Solving a QUBO problem means finding a binary vector \( |\mathbf{x}^* \rangle \) that minimizes \( f_Q \),
\begin{equation}
|\mathbf{x}^* \rangle = \arg \min f_Q(\mathbf{x}).
\end{equation}
QUBO problems are good candidates for quantum algorithms because there is a simple mapping that enables them to be executed on a quantum computer. With the goal of linking QUBO to QAOA, we can rewrite the objective function in Equation \eqref{eq:qubo}  as a QUBO Hamiltonian
\begin{equation} \label{eq:ham_qubo}
H_{\text{QUBO}} = \sum_{i\neq j} Q_{ij} x_i x_j + \sum_{i=1}^n (Q_{ii}+c_i) x_i,
\end{equation}
where the second term has been simplified using the fact that, for binary variables, \( x_i^2 = x_i \). 
Then, a simple linear transformation of the form \( Z \gets 2\mathbf{x} - \mathbf{1} \), where $\mathbf{1} = {\underbrace{[1, 1, \dots, 1]}_{n}}^\top$, relates the two problems, with matrix $J$ occuring from this simple linear transformation. Without loss of generality we therefore proceed to map the problem to a quantum computer by using the following Hamiltonian
\begin{equation}
H_C = \sum_{i\neq j} J_{ij} Z_i Z_j + \sum_{i=1}^n J_{ii} Z_i.
\end{equation}
The Hamiltonian \( H_C \) is named the cost Hamiltonian \cite{farhi2014quantum}.

The QAOA approach also involves initializing a quantum state \( |\psi_0\rangle \), typically an equal superposition of all computational basis states, as follows
\begin{equation}
    |\psi_0\rangle = |+\rangle^{\otimes n} = \frac{1}{\sqrt{2^n}} \sum_{\mathbf{x} \in \{0, 1\}^n} |\mathbf{x}\rangle.
\end{equation}
The algorithm then applies a series of alternating unitary operators; a cost unitary \( U_C(\gamma) = e^{-\mathfrak{i} \gamma H_C} \) and a mixer unitary \( U_M(\beta) = e^{-\mathfrak{i} \beta H_M} \), where, \( \gamma, \beta \) are real-valued parameters that control the evolution of the system.  The cost unitary \( U_C(\gamma) = e^{-\mathfrak{i} \gamma H_C} \) applies a phase proportional to the cost function, effectively encoding the objective into the quantum state's phase. The term \( H_M = \sum_{j=1}^n X_j \) is the \textit{mixer Hamiltonian}, with \( X_j\) being the Pauli-X operator acting on qubit \( j \). It is noted that the mixer unitary \( U_M(\beta) = e^{-\mathfrak{i} \beta H_M} \) encourages exploration of the solution space by flipping the state of each qubit with amplitude \( \beta \). Therefore, the time evolution is driven by the cost Hamiltonian, while ``hopping" between states is enabled by the mixer unitary. The role of the mixer unitary is crucial to ensure that the system’s dynamics remain non-trivial. If the system evolved solely under the influence of the cost Hamiltonian, the energy, or cost,  associated with it would remain constant \cite{abbas2023quantum}. Thus, varying $\gamma$ would have no effect on the cost, making optimization impossible. By selecting a mixing Hamiltonian that does not commute with the cost Hamiltonian, the cost is no longer conserved, allowing for the identification of a minimum by adjusting $\gamma,\beta$. Then, for an integer \( p \), which controls the circuit depth, the QAOA state after \( p \) layers is then defined as
\begin{equation}
    |\psi(\boldsymbol{\gamma}, \boldsymbol{\beta})\rangle = \prod_{k=1}^{p} U_M(\beta_k) U_C(\gamma_k) |\psi_0\rangle,
\end{equation}
where \( \boldsymbol{\gamma} = (\gamma_1, \ldots, \gamma_p) \) and \( \boldsymbol{\beta} = (\beta_1, \ldots, \beta_p) \) are vectors of parameters. 
\textcolor{black}{Despite being a heuristic, the motivation around
QAOA stems from the fact that, even
for finite values of $p$, there are problem instances for which
practical implementations of QAOA have been shown to
achieve competitive or improved performance against classical
benchmarks \cite{zhou2020quantum}. Discovering such problem instances is of practical interest.}

Then, the objective is to minimize the expectation value of the cost Hamiltonian \( H_C \) in the state \( |\psi(\boldsymbol{\gamma}, \boldsymbol{\beta})\rangle \), which is given by
\begin{equation}
    F(\boldsymbol{\gamma}, \boldsymbol{\beta}) = \langle \psi(\boldsymbol{\gamma}, \boldsymbol{\beta}) | H_C | \psi(\boldsymbol{\gamma}, \boldsymbol{\beta}) \rangle.
\end{equation}
This expectation value is evaluated on a quantum computer, and a classical optimizer iteratively updates \( \boldsymbol{\gamma} \) and \( \boldsymbol{\beta} \) to approximate the optimal \( \boldsymbol{\gamma}^* \) and \( \boldsymbol{\beta}^* \) that satisfies 
\begin{equation}
    \boldsymbol{\gamma}^*, \boldsymbol{\beta}^* = \arg \min F(\boldsymbol{\gamma}, \boldsymbol{\beta}).
\end{equation}
For instance, the Constrained Optimization by Linear Approximations (COBYLA) \cite{powell2007view} is a gradient-free optimizer that iteratively adjusts the parameters by building simple linear models that approximate the objective function, based on previous sample evaluations. \textcolor{black}{At each step, the COBYLA proposes a new set of parameters, the quantum circuit is executed with these parameters to estimate the expected value, and then uses this information to refine its next guess. By continuously updating the parameters in this manner, COBYLA efficiently navigates the parameter space to find the values that achieve the lowest expected cost, thereby guiding QAOA to approximate the optimal solution for the given optimization problem. After the optimal parameters \( \boldsymbol{\gamma}^* \) and \( \boldsymbol{\beta}^* \) have been found, the quantum state \( |\psi(\boldsymbol{\gamma}^*, \boldsymbol{\beta}^*)\rangle \) is measured multiple times. Each measurement collapses the state into a specific binary vector \( \mathbf{x}^* \) with a probability determined by the amplitude of that configuration in \( |\psi(\boldsymbol{\gamma}^*, \boldsymbol{\beta}^*)\rangle \). Since the optimal parameters \( \boldsymbol{\gamma}^* \) and \( \boldsymbol{\beta}^* \) have been chosen to minimize the expected value of the Hamiltonian $H_C$ in (9), measuring the state \( |\psi(\boldsymbol{\gamma}^*, \boldsymbol{\beta}^*)\rangle \) repeatedly yields a vector \( \mathbf{x}^* \), which is optimal for the original problem of (2) with high probability.} \textcolor{black}{This implies that any classical optimizer, including COBYLA, can fail to find the optimal set of parameters $\boldsymbol{\gamma}^*, \boldsymbol{\beta}^*$. A primary cause for this is the complex, non-convex landscape of the QAOA method, which will be further discussed in section VII.}


\subsection{Warm-start QAOA (WS-QAOA)}
The solutions of a continuous-valued relaxation of a QUBO problem can be used to initialize the corresponding quantum-classical hybrid algorithm, a process known as warm-start \cite{egger2021warm}. In the simplest variant of WS-QAOA, the initial equal superposition state $\lvert + \rangle^{\otimes n}$ is replaced with the state
\begin{equation}
\lvert \psi_0 \rangle = \bigotimes_{i=1}^{n} R_{Y_i}(y_i) \lvert 0 \rangle,
\end{equation}
which corresponds to the optimal solution $\mathbf{c}^*$ of the relaxed version of problem (3). Here, $R_{Y_i}(y_i)$ is a rotation around the $Y$-axis of the $i$-th qubit with angle $y_i = 2 \arcsin(\sqrt{c_i^*})$ and $c_i^* \in [0,1]$ is the $i$-th coordinate of the relaxed problem's solution. The mixer Hamiltonian $H_M$ is also replaced with $H_M^{\text{ws}} = \sum_{i=1}^{n} H_{M,i}^{\text{ws}}$, where
\begin{equation}
H_{M,i}^{\text{ws}} = \begin{pmatrix}
2c_i^* - 1 & -2\sqrt{c_i^*(1-c_i^*)} \\
-2\sqrt{c_i^*(1-c_i^*)} & 1 - 2c_i^*
\end{pmatrix}
\end{equation}
and has $R_{Y_i}(y_i)\lvert 0 \rangle$ as the ground state with eigenvalue $-1$. 
Therefore, WS-QAOA applies at the $k$-th layer a mixing gate defined by the time-evolved mixing Hamiltonian $\exp(-i\beta_k H_M^{\text{ws}})$. Moreover, since $H_{M,i}^{\text{ws}} = -\sin(y_i)X - \cos(y_i)Z$, the time-evolved mixing Hamiltonian represents a rotation around the axis $\vec{n} = [-\sin(y_i), 0, -\cos(y_i)]$ on the Bloch sphere of the $i$-th qubit, and can be implemented using the single-qubit rotations $R_{Y_i}(y_i)R_{Z_i}(-2\beta_k)R_{Y_i}(-y_i)$. \textcolor{black}{Thus, the WS-QAOA procedure mirrors that of QAOA, differing only in the initial state and the mixer Hamiltonian used.}

\textcolor{black}{ As a final note on QAOA, it should be mentioned that the performance of QAOA in practice can be strongly affected by quantum noise such as bit and phase flips, decoherence and etc. These issues are not unique to QAOA but are common to essentially all near-term quantum computing algorithms, where noise and imperfect operations reduce the fidelity of computation and hence the quality of results. Recent works have explored error mitigation strategies \cite{hanzo}, which can partially alleviate these effects on current devices. In the long term, improved quantum error correction will be required to achieve reliable large-scale implementations \cite{hanzo}.}

\section{System Model}
We consider a MIMO system consisting of $N_\mathrm{T}$ and $N_\mathrm{R}$ transmit and receive antennas, respectively, \textcolor{black}{with metamaterial insulators in order to neglect the mutual coupling \cite{coupling}. Nonetheless, this work can be extended in the case of mutual coupling as well.} Let $H \in \mathbb{C}^{N_\mathrm{R} \times N_\mathrm{T}}$ denote the MIMO channel matrix, where entries correspond to the channels between the transmit and receive antennas, i.e., $h_{i,j}$ is the channel coefficient between the $j$-th transmit antenna and the $i$-th receive antenna. The MIMO system operates in the \textit{beamforming mode}  to maximize the received SNR and thus the pre/post-processing vectors have to be adjusted accordingly. 

To reduce the implementation complexity and power consumption, both the transmitter and the receiver are equipped with \(b\)-bit resolution phase shifters. Let \(\mathbf{f} \in \mathcal{S}^{N_\mathrm{T} \times 1}\) and \(\mathbf{g} \in \mathcal{S}^{N_\mathrm{R} \times 1}\) denote the unnormalized pre-coding and post-coding vectors, respectively. The set $\mathcal{S}$ is the set containing the available quantized phase shifts. 
The normalized pre/post-coding vectors are then given by $\mathbf{f} / \|\mathbf{f}\| = \mathbf{f} / \sqrt{N_\mathrm{T}}$ and $\mathbf{g} / \|\mathbf{g}\| = \mathbf{g} / \sqrt{N_\mathrm{R}}$, respectively. We assume a perfect channel state information at both the transmitter and the receiver. The received SNR is given by
\begin{equation}
    \rho(\mathbf{g}, \mathbf{f}) = \frac{P \left| \mathbf{g}^{\dagger} H \mathbf{f} \right|^2}{N_\mathrm{T} N_\mathrm{R} \sigma^2},
\end{equation}
where $\sigma^2$ is the variance of the additive white Gaussian noise, $P$ is the transmit power, while the terms $N_\mathrm{T}$ and $N_\mathrm{R}$ in the denominator are due to the power normalization of the pre/post-coding vectors. Since the objective of the MIMO system is to maximize the received SNR, we introduce the following design problem
\begin{equation}\label{eq:opt_original}
    \max_{\mathbf{f} \in \mathcal{S}^{N_\mathrm{T} \times 1}, \, \mathbf{g} \in \mathcal{S}^{N_\mathrm{R} \times 1}} \left| \mathbf{g}^{\dagger} H \mathbf{f} \right|^2.
\end{equation}
\textcolor{black}{Due to the binary nature of the analogue pre/post-coding vectors, the above optimization problem is combinatorial and NP-hard; there is no known polynomial-time algorithm, which  is guaranteed to obtain the global optimum.}

\section{QAOA-based beamforming optimization, $b = 2$}
In this section we present how to transform problem (13), for $b=2$, in quadratic form, which can be directly solved using QAOA. For $b = 2$, we have \( \mathcal{S} \!=\! \{e^{\mathfrak{i}\pi/4}, e^{\mathfrak{i}3\pi/4}, e^{\mathfrak{i}5\pi/4}, e^{\mathfrak{i}7\pi/4}\} = \{1 + \mathfrak{i}, -1+\mathfrak{i}, -1-\mathfrak{i}, 1-\mathfrak{i}\}\). From (13), we observe that the function we need to optimize, $\left| \mathbf{g}^{\dagger} H \mathbf{f} \right|^2 = \mathbf{f}^{\dagger} H^{\dagger}\mathbf{g} \mathbf{g}^{\dagger} H \mathbf{f}$, is not in QUBO form as defined in \eqref{eq:qubo}. Also, $\mathbf{f}\in\mathcal{S}^{N_\mathrm{T}}$ and $\mathbf{g}\in\mathcal{S}^{N_\mathrm{R}}$ instead of $\{0,1\}^{N_\mathrm{T}}$ and $\{0,1\}^{N_\mathrm{R}}$ in respect. To transform problem (13) into a form which can be solved via QAOA, the AO method will be~employed.
\subsubsection{Optimization of $\mathbf{f}$} When optimizing with respect to (w.r.t.) $\mathbf{f}$, we fix the variable $\mathbf{g}$ to a constant value of choice. Then, we also set the new matrix $ A = \mathbf{g}^{\dagger} H$ which results in the following optimization problem
\begin{equation}\label{eq:opt_original}
    \max_{\mathbf{f} \in \mathcal{S}^{N_\mathrm{T} \times 1}} \left|  A \mathbf{f} \right|^2 =  \mathbf{f}^{\dagger} A^{\dagger} A \mathbf{f},
\end{equation}
which is in a quadratic form. Nonetheless, variable $\mathbf{f}$ is still not binary but a complex one in the set $\mathcal{S}$. As such we define the following quantities
\begin{equation}
    \mathbf{f}_\mathrm{R} = \operatorname{Re}\{\mathbf{f}\}, \, \mathbf{f}_\mathrm{I} = \operatorname{Im}\{\mathbf{f}\},\,  A_\mathrm{R} = \operatorname{Re}\{ A\}, \,  A_\mathrm{I} = \operatorname{Im}\{ A\},
\end{equation}
where, with a slight abuse of notation, the operators $\operatorname{Re}\{\cdot\}$ and $\operatorname{Im}\{\cdot\}$ denote the element-wise real and the imaginary part respectively, of a vector or a matrix with complex-valued entries. Then, the following holds
\begin{equation}
\begin{aligned}
     A\mathbf{f} &= ( A_\mathrm{R} + \mathfrak{i}  A_\mathrm{I})(\mathbf{f}_\mathrm{R} + \mathfrak{i} \mathbf{f}_\mathrm{I}) \\
    & =  A_\mathrm{R} \mathbf{f}_\mathrm{R} -  A_\mathrm{I} \mathbf{f}_\mathrm{I} + \mathfrak{i} ( A_\mathrm{R} \mathbf{f}_\mathrm{I} +  A_\mathrm{I} \mathbf{f}_\mathrm{R}).
\end{aligned}
\end{equation}
Thus, the real and imaginary parts of \(  A\mathbf{f} \) are:
\begin{align}
\operatorname{Re}\{ A\mathbf{f}\} =  A_\mathrm{R} \mathbf{f}_\mathrm{R} -  A_\mathrm{I} \mathbf{f}_\mathrm{I}, \,\, \operatorname{Im}\{ A\mathbf{f}\} =  A_\mathrm{R} \mathbf{f}_\mathrm{I} +  A_\mathrm{I} \mathbf{f}_\mathrm{R}.
\end{align}
Writing the product of \( | A\mathbf{f}|^2 \)  in terms of real and imaginary parts, we have
\begin{equation}\label{eq:qf_qubo_like}
\begin{aligned}    
| A\mathbf{f}|^2 &= \left( A_\mathrm{R} \mathbf{f}_\mathrm{R} -  A_\mathrm{I} \mathbf{f}_\mathrm{I}\right)^2 + \left( A_\mathrm{R} \mathbf{f}_\mathrm{I} +  A_\mathrm{I} \mathbf{f}_\mathrm{R}\right)^2 \\
&= ( A_\mathrm{R} \mathbf{f}_\mathrm{R} -  A_\mathrm{I} \mathbf{f}_\mathrm{I})^\top ( A_\mathrm{R} \mathbf{f}_\mathrm{R} -  A_\mathrm{I} \mathbf{f}_\mathrm{I})  \\
&\quad\, + ( A_\mathrm{R} \mathbf{f}_\mathrm{I} +  A_\mathrm{I} \mathbf{f}_\mathrm{R})^\top ( A_\mathrm{R} \mathbf{f}_\mathrm{I} +  A_\mathrm{I} \mathbf{f}_\mathrm{R}) \\
 &= \mathbf{f}_\mathrm{R}^\top\!\left( A_\mathrm{R}^\top A_\mathrm{R} +  A_\mathrm{I}^\top A_\mathrm{I}\right)\!\mathbf{f}_\mathrm{R} + \mathbf{f}_\mathrm{I}^\top\!\left( A_\mathrm{R}^\top A_\mathrm{R} +  A_\mathrm{I}^\top A_\mathrm{I}\right)\!\mathbf{f}_\mathrm{I}.
\end{aligned}
\end{equation} 
We note that the variables $\mathbf{f}_{\mathrm{R}}$ and $\mathbf{f}_{\mathrm{I}}$ are in $\{-1,1\}$, thus by using the transformations $\mathbf{f}_\mathrm{R} = 2\mathbf{x}_{\mathrm{R}}-1$, $\mathbf{f}_\mathrm{I} = 2\mathbf{x}_{\mathrm{I}}-1$, 
and by setting 
\begin{equation}
  Q = 4\begin{bmatrix} A_{\mathrm{R}}^\top A_{\mathrm{R}} +  A_{\mathrm{I}}^\top A_{\mathrm{I}} & 2 A_{\mathrm{I}}^\top A_{\mathrm{R}} \\ -2 A_{\mathrm{I}}^\top A_{\mathrm{R}} &  A_{\mathrm{R}}^\top A_{\mathrm{R}} +  A_{\mathrm{I}}^\top A_{\mathrm{I}} \end{bmatrix},
\end{equation}
the following holds
\begin{equation}\label{eq:qubo_f}
| Af|^2 = \mathbf{x}^\top   Q \mathbf{x} - \mathbf{x}^\top   Q \mathbf{1} + \mathbf{1}^\top   Q \mathbf{1},
\end{equation}
where $\mathbf{x} = \left[\mathbf{x}_{\mathrm{R}}\,\,\mathbf{x}_{\mathrm{I}}\right]^\top$ and $\mathbf{1} = {\underbrace{[1, 1, \dots, 1]}_{N_\mathrm{T}}}^\top$. By ignoring the constant term, \eqref{eq:qubo_f} is in QUBO form according to \eqref{eq:qubo} and can be solved by following the analysis of section II.A.
\begin{algorithm}
\caption{QAOA-based beamforming optimization, $b=2$}
\textbf{Input:} Matrix $H \in \mathbb{C}^{N_\mathrm{T}\times N_\mathrm{R}}$, initial complex vectors $g^0\!\in\! \mathbb{S}^{\mathrm{N}\mathrm{r}}$, maximum number of iterations $K$ 
\begin{algorithmic}[1]
\For{$k = 1, 2, \dots, K$}
        \If{$k = 1$}
            \State $\mathbf{g}^{*} \gets \mathbf{g}^{0}$
        \EndIf
        \State Set $ A = \mathbf{g}^{*\dagger}H$, $ A =  A^\dagger A$, $  Q$ following (18), and convert to QUBO form according to section III
        \State \textbf{[QAOA]} Solve for $\mathbf{f}^*$ using QAOA: $\mathbf{f}^* = \mathbf{x}_\mathrm{R}^*+i\mathbf{x}_\mathrm{I}^*$, where $\mathbf{x}^* = \arg \min_\mathbf{x} -\mathbf{x}^\top   Q \mathbf{x} + \mathbf{x}^\top   Q \mathbf{1}$
        \State Convert $\mathbf{f}^*$ to spin vector using $\mathbf{f}^{*} \gets 2\mathbf{f}^*-\mathbf{1}$
        
        \State Set $ A = (H\mathbf{f}^{*})^\dagger$, $\mathbf{g}\gets\mathbf{g}^\dagger$ $ A =  A^\dagger A$, $  Q$ following (18), and convert to QUBO form according to section III
        \State \textbf{[QAOA]} Solve for $\mathbf{g}^*$ using QAOA: $\mathbf{g}^* = \mathbf{x}_\mathrm{R}^*+i\mathbf{x}_\mathrm{I}^*$, where $\mathbf{x}^* = \arg \min_\mathbf{x} -\mathbf{x}^\top   Q \mathbf{x} + \mathbf{x}^\top   Q \mathbf{1}$
        \State Convert $\mathbf{g}^*$ to spin vector using $\mathbf{g}^{*} \gets 2\mathbf{g}^*-\mathbf{1}$ and then $\mathbf{g}\gets\mathbf{g}^\dagger$
        \State Compute $\rho_{\text{new}}^k = \rho(\mathbf{f}^*, \mathbf{g}^*)$
        \If{$\rho_{\text{new}}^k < \rho_{\text{new}}^{k-1}$}
            \State \textbf{break}  
        \EndIf
\EndFor
\State \textbf{Output:} Optimized $(\mathbf{f}^*, \mathbf{g}^*) = \arg \max_{\mathbf{f}, \mathbf{g}} \rho(\mathbf{f}, \mathbf{g})$
\end{algorithmic}
\end{algorithm}
\subsubsection{Optimization of $\mathbf{g}$} A similar method is employed to optimize variable $\mathbf{g}$. Specifically, we define the matrix $ A = H \mathbf{f}$ and the considered optimization problem is then given as
\begin{equation}\label{eq:opt_g}
    \max_{\mathbf{g} \in \mathcal{S}^{N_\mathrm{R} \times 1}} \left| \mathbf{g}  A \right|^2 =   A\mathbf{g}^\dagger \mathbf{g}  A^\dagger = \mathbf{g}  A A^\dagger \mathbf{g}^\dagger.
\end{equation}
This problem is in quadratic form, but to transform it as in \eqref{eq:qubo}, with a slight abuse of notation, we have to set $\mathbf{g} \gets \mathbf{g}^\dagger$ and $ A \gets  A^\dagger$. Then, the problem to optimize is given as 
\begin{equation}\label{eq:opt_g}
    \max_{\mathbf{g} \in \mathcal{S}^{N_\mathrm{R} \times 1}} \mathbf{g}^\dagger  A^\dagger A \mathbf{g},
\end{equation}
which is similar to \eqref{eq:qubo}. From there, the analysis to transform it to a QUBO problem, follows that of the previous subsection.
Algorithm 1 provides a comprehensive overview of the iterative QAOA procedure which solves the low-resolution beamforming optimization problem with $b=2$. \textcolor{black}{We note that the notation $\boldsymbol{f}^*,\boldsymbol{g}^*$ denote the best solution found, which may not necessarily be optimal. The performance of Alg. 1 will be numerically evaluated.} 

\section{QAOA-based $b$-bit beamforming optimization}
In this section we aim to generalize this analysis of the previous section for an arbitrary number of quantization bits $b$. For any $b$, the available phase shifting angles are given by 
\begin{equation} \label{eq:phi_orig}
    \phi_i = {\frac{(2i+1)\pi}{2^b}},\,\,\forall i = 0,1,...,2^b-1,
\end{equation}
thus $\mathcal{S} = \{e^{\mathfrak{i}\phi_i},\,\,\forall i = 0,1,...,2^b-1\}$. To utilize off-the-self QAOA implementations, we need to express the $b$-bit beamforming optimization problem into a binary optimization form. A straightforward approach, based on the analysis of the previous section, is to define the pre- and post-coding vectors as follows 
\begin{equation}
    \mathbf{f} = \begin{pmatrix}\sum_{i = 0}^{2^b-1}x^1_ie^{\mathfrak{i}\phi_i} \\ \vdots \\ \sum_{i = 0}^{2^b-1}x^{N_\mathrm{T}}_ie^{\mathfrak{i}\phi_i} \end{pmatrix}, \,\,\, 
    \mathbf{g} = \begin{pmatrix}\sum_{i = 0}^{2^b-1}x^1_ie^{\mathfrak{i}\phi_i} \\ \vdots \\ \sum_{i = 0}^{2^b-1}x^{N_\mathrm{R}}_ie^{\mathfrak{i}\phi_i} \end{pmatrix},
\end{equation}
where $x^j_i,\forall i,j$ are binary variables. For now, let us assume that $\mathbf{g}$ is fixed. Then, the $b$-bit beamforming optimization problem, is given as follows 
\begin{equation}\label{eq:opt_one_hot}
    \begin{aligned}
     \max_{\mathbf{x}} & \begin{pmatrix}\sum_{i = 0}^{2^b-1}x^1_ie^{\mathfrak{i}\phi_i} \\ \vdots \\ \sum_{i = 0}^{2^b-1}x^{N_\mathrm{T}}_ie^{\mathfrak{i}\phi_i} \end{pmatrix}^{\dagger} A^{\dagger} A \begin{pmatrix}\sum_{i = 0}^{2^b-1}x^1_ie^{\mathfrak{i}\phi_i} \\ \vdots \\ \sum_{i = 0}^{2^b-1}x^{N_\mathrm{T}}_ie^{\mathfrak{i}\phi_i} \end{pmatrix} \\
    \text{s.t.} & \sum_{i = 0} ^{2^b-1} x^j_i = 1,  \,\, j = 1,...,N_\mathrm{T},
    \end{aligned}
\end{equation}
where $\mathbf{x}$ contains all $x^j_i, \forall j,i$. The problem in this form can be solved using QAOA, however there are two notes to make. First, this encoding results to ``additional" basis states in the Hilbert space that do not correspond to valid solutions, since more than one qubit might accidentally equal one, or all of them could equal zero. As such an equality constraint was added which aims to enforce a valid solution. Then, the constraint term needs to be added as a penalty term in the problem's Hamiltonian. But, penalty terms must be carefully tuned; too large a penalty can hamper the QAOA dynamics, whereas too small a penalty may not reliably enforce the constraint. Second, this optimization problem requires a total number of $2^bN_\mathrm{T}$ of qubits, which increases exponentially with the number of quantization level $b$. Keeping the required number of qubits minimal is a critical factor for the practical implementation of any quantum algorithm in the NISQ era.

To avoid both drawbacks, we will propose a different encoding for this problem. In \eqref{eq:phi_orig} we can omit the term $\pi/2^b$, since due to the quadratic form of the SNR, given in (12), the phase shift term $e^{\mathfrak{i}\pi/2^b}$ is always being multiplied by the phase shift term  $e^{-\mathfrak{i}\pi/2^b}$, thus being canceled out. As such, we consider 
\begin{equation}
    \phi_i = {\frac{2i\pi}{2^b}},\,\,\forall i = 0,1,...,2^b-1.
\end{equation}
Then, the following lemma holds:
\begin{lemma} For any $b\in \mathbb{N}$, there is a subset of exactly \(b\) angles, $
  \{\theta_0,\!\theta_1,\!\dots,\!\theta_{b-1}\}\!\subset\!\{\phi_0,\dots,\phi_{2^b-1}\},$ such that \emph{every} \(\phi_i\) can be written as
\(
  \phi_i 
  =
  \sum_{j=0}^{b-1} x_j\,\theta_j\), where each $x_j\in \{0,1\}$.
Equivalently, the map 
\(\,(\phi_0,\dots,\phi_{2^b-1}) \,\mapsto\, \sum_{j=0}^{b-1}x_j\,\theta_j\)
holds.
\end{lemma}
\begin{IEEEproof} The proof is given in Appendix A. \end{IEEEproof}

Following this lemma, we define the pre- and the post-coding vectors as 
\begin{equation}
    \mathbf{f} = \begin{pmatrix}e^{\mathfrak{i}\sum_{i = 0}^{b-1}x^1_i\theta_i} \\ \vdots \\ e^{\mathfrak{i}\sum_{i = 0}^{b-1}x^{N_\mathrm{T}}_i\theta_i} \end{pmatrix}, \,\,\, 
    \mathbf{g} = \begin{pmatrix}e^{\mathfrak{i}\sum_{i = 0}^{b-1}x^1_i\theta_i} \\ \vdots \\ e^{\mathfrak{i}\sum_{i = 0}^{b-1}x^{N_\mathrm{T}}_i\theta_i} \end{pmatrix},
\end{equation}
respectively. We note that these vectors correspond only to $bN_\mathrm{T}$ and $bN_\mathrm{R}$ qubits. For now, let us assume that $\mathbf{g}$ is fixed, then the optimization problem, w.r.t. $\mathbf{f}$, is given by
\begin{equation}\label{eq:opt_f_hubo}
    \begin{aligned}
     \max_{\mathbf{x}} & \begin{pmatrix}e^{\mathfrak{i}\sum_{i = 0}^{b-1}x^1_i\theta_i} \\ \vdots \\ e^{\mathfrak{i}\sum_{i = 0}^{b-1}x^{N_\mathrm{T}}_i\theta_i} \end{pmatrix}^{\dagger} A^{\dagger} A \begin{pmatrix}e^{\mathfrak{i}\sum_{i = 0}^{b-1}x^1_i\theta_i} \\ \vdots \\ e^{\mathfrak{i}\sum_{i = 0}^{b-1}x^{N_\mathrm{T}}_i\theta_i} \end{pmatrix},
    \end{aligned}
\end{equation}
where $A$ is as defined in section IV. We can immediately observe that this problem is unconstrained, but also that in the general case of $b>2$, this problem is not in standard QUBO form, as it was defined in section II, since all optimization variables are exponentiated. Nonetheless, this problem can be still solved using a QAOA-based approach, as it will be shown in the following subsections. However, unlike the previous case, we now have to explicitly calculate the problem's Hamiltonian.
\subsection{An illustrative example}
\begin{figure}[h!]
    \centering
    \resizebox{.8\columnwidth}{!}{ 
        \input{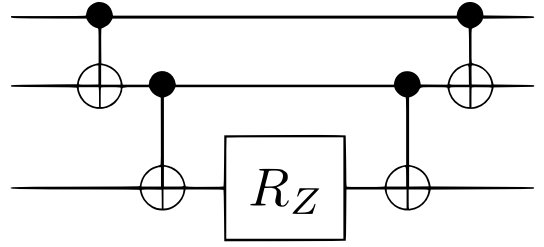} 
    }
    \caption{The 3-qubit rotation gate.}
    \label{fig:Rzgate}
\end{figure}

Before presenting the general analysis for this problem, let us examine the case of $N_\mathrm{T} = 2$ and $b = 1$. In addition, since $A^{\dagger}A$ is Hermitian, let's assume that  $A^{\dagger}A = \begin{pmatrix} a & c \\ c^{\dagger} & d \end{pmatrix}$. Then, after some simple algebraic manipulations and omitting constant terms, the objective function is given as 
\begin{equation}
\begin{aligned}
    \rho(\mathbf{x}) = c^{\dagger}e^{\mathfrak{i}(\theta_1x_1 - \theta_2x_2)} + ce^{\mathfrak{i}(\theta_2x_2 - \theta_1x_1)} 
\end{aligned}
\end{equation}
By using $Z \gets 2\mathbf{x} - \mathbf{1}$, the cost Hamiltonian corresponding to this objective is given as follows
\begin{equation}
    H_C = c^{\dagger}e^{\mathfrak{i}(\theta_1Z_1 - \theta_2Z_2)} + ce^{\mathfrak{i}(\theta_2Z_2 - \theta_1Z_1)}.
\end{equation}
We observe that the Hamiltonian is Hermitian, which is necessary for the Hamiltonian of the problem to be feasible. By considering that the rotation operator about the z-axis is given as 
\begin{equation} \label{eq:R_z}
R_Z(\theta) \;\equiv\; e^{-\,\mathfrak{i}\,\frac{\theta}{2}Z}
\;=\;
\cos \biggl(\frac{\theta}{2}\biggr)\,I 
\;-\; \mathfrak{i}\,\sin\!\biggl(\frac{\theta}{2}\biggr)\,Z,
\end{equation}
the Hamiltonian is equivalently given as 
\begin{equation}
    H_C = c^{\dagger}R_{Z_1}(-2\theta_1)\otimes R_{Z_2}(2\theta_2) + cR_{Z_1}(2\theta_1)\otimes R_{Z_2}(-2\theta_2),
\end{equation}
where $R_{Z_i}(-2\theta_1)$ means that the rotation operator acts on the i-th qubit. As discussed in Section II, every quantum circuit, without the effects of noise, can be described by a unitary transformation and for this unitary transformation $U$ it holds that $U = e ^{-\mathfrak{i}Ht}$. Thus, the quantum circuit describing the cost Hamiltonian is given by the operator,
\begin{equation} \label{eq:U_c}
    U_C = e^{-\mathfrak{i}\left(c^{\dagger}e^{\mathfrak{i}(\theta_1Z_1 - \theta_2Z_2)} + ce^{\mathfrak{i}(\theta_2Z_2 - \theta_1Z_1)}\right)},
\end{equation}
which is not straightforward how to implement. Therefore, by using \eqref{eq:R_z}, we have that
\begin{equation} \label{eq:rtensorr}
\begin{aligned}
& R_{Z_1}(\theta_1) \otimes R_{Z_2}(\theta_2) = \\
& \cos\left(\frac{\theta_1}{2}\right)\cos\left(\frac{\theta_2}{2}\right) I \otimes I 
- \mathfrak{i}\sin\left(\frac{\theta_1}{2}\right)\cos\left(\frac{\theta_2}{2}\right) Z \otimes I \\
& - \sin\left(\frac{\theta_1}{2}\right)\sin\left(\frac{\theta_2}{2}\right) Z \otimes Z 
- \mathfrak{i}\cos\left(\frac{\theta_1}{2}\right)\sin\left(\frac{\theta_2}{2}\right) I \otimes Z.
\end{aligned}
\end{equation}
Then, the cost Hamiltonian of (33) is given as
\begin{equation}
\begin{aligned}
    H_C &= 2c_I\sin{\theta_1}\cos{\theta_2}Z\otimes I \\
    &+ 2c_R\sin{\theta_1}\sin{\theta_2}Z\otimes Z - 2c_I\cos{\theta_1}\sin{\theta_2} I\otimes Z,
\end{aligned}
\end{equation}
where $c_I = \mathrm{Im}\left\{c\right\}$ and $c_R = \mathrm{Re}\left\{c\right\}$, while the term containing only $I \otimes I$ has been omitted since it does not affect the optimal solution of the problem. Since the Hamiltonian has been decomposed to terms containing the Pauli matrices $I,Z$ it is easy to create the quantum circuit associated with the Hamiltonian. Let us denote the 2-qubit CNOT gate as follows
\begin{equation}
    \text{CNOT} = \ket{0}\bra{0} \otimes I + \ket{1}\bra{1} \otimes X.
\end{equation}
By taking this formula into account, we can show that the circuit which implements the term \( e^{\mathfrak{i} Z \otimes Z t} =  \cos{t}I+\mathfrak{i}\sin{t}Z\otimes Z\), where $t$ any real number, is the following
\begin{equation}
\begin{aligned}
&\text{CNOT} \, (I \otimes e^{\mathfrak{i} Z t}) \, \text{CNOT} = \left[ \ket{0}\bra{0} \otimes I + \ket{1}\bra{1} \otimes X \right] \\
&\cdot\left[ \cos(t) I \otimes I + \mathfrak{i} \sin(t) I \otimes Z \right] \cdot \left[ \ket{0}\bra{0} \otimes I + \ket{1}\bra{1} \otimes X \right] \\
& = \cos{t}I+\mathfrak{i}\sin{t}Z\otimes Z.
\end{aligned}
\end{equation}
We note that this formula can be generalized for any \( e^{\mathfrak{i} Z^{\otimes n}}\) and the 3-qubit case is shown in Fig. \ref{fig:Rzgate}. Then, taking the exponential of $H_C$ as in  \eqref{eq:U_c}, we conclude that the cost unitary is given as
\begin{equation}
\begin{aligned}
    U_C = & \left[ R_{Z}\left(2\gamma c_I\sin{\theta_1}\cos{\theta_2}\right) \otimes I \right] \\
    & \cdot\left[\text{CNOT} \, \left( I \otimes R_Z\left(2\gamma c_R\sin{\theta_1}\sin{\theta_2}\right) \right) \, \text{CNOT} \right] \\
    & \cdot \left[ I\otimes R_{Z}\left(-2\gamma c_I\cos{\theta_1}\sin{\theta_2}\right) \right],
\end{aligned}
\end{equation}
which can be used to realize the QAOA circuit, according to section II and Algorithm 2. From this illustrative example, it is obvious that the phase shifts used in beamforming can be directly mapped to rotation gates in a quantum circuit.
\subsection{The General case}
In the case of $b\geq 1$ and $N_\mathrm{T},N_\mathrm{R} \geq 2$ the main issue is to find a way to calculate the general form of the Hamiltonian, based only on the Pauli matrices $Z,I$. 
Let us define $Q = A^{\dagger}A$ and $Q_{ij}$ to be the element of the $i$-th raw and the $j$-th row. Since $Q$ is a Hermitian matrix, then $Q_{ij} = Q^{\dagger}_{ij}$. As such, the objective of problem \eqref{eq:opt_f_hubo} is given as 
\begin{equation}
    \rho(\mathbf{x}) = \sum_{k = 1}^{N_\mathrm{T}}\sum_{\substack{j=1 \\ j \neq k}}^{N_\mathrm{T}} Q_{kj} e^{\mathfrak{i}\left(-\sum_{i = 0}^{b-1}x^k_i\theta_i + \sum_{i = 0}^{b-1}x^j_i\theta_i\right)}.
\end{equation}
Due to the Hermittian symmetry of $Q$ and the quadratic nature of the objective function, it is guaranteed that the Hamiltonian corresponding to this function will always be Hermitian. Moreover, for ease of notation, we introduce the following tranformation
\begin{equation}
\begin{aligned}
    &  x^k_i \mapsfrom  \frac{Z_{(k-1)b+i} + 1}{2}.
\end{aligned}
\end{equation}
Based on (39) and the previous transformation, the $k,j$-th term of the Hamiltonian is then given as 
\begin{equation*}
    Q_{kj} e^{\mathfrak{i}\left(-\sum_{i = 0}^{b-1} Z_{(k-1)b+i} \theta_i + \sum_{i = 0}^{b-1} Z_{(j-1)b+i} \theta_i\right)},
\end{equation*}
\textcolor{black}{where the sum $-\sum_{i}Z_{(k-1)b+i}\,\theta_i$ applies a negative phase shift of $\theta_i$ to each of the $b$ qubits in block $k$, while the sum $+\sum_{i}Z_{(j-1)b+i}\,\theta_i$ applies a positive phase shift of $\theta_i$ to each qubit in block $j$}. In addition, global phase factors have been omitted. Moreover, for any Pauli operator $P$ the following holds 
\begin{equation}
\begin{aligned}    
e^{\mathfrak{i} I \otimes P \otimes I t} &= \cos(t) I \otimes I \otimes I + \mathfrak{i} \sin(t) I \otimes P \otimes I \\
&= I \otimes (\cos(t) I + \mathfrak{i} \sin(t) P) \otimes I = I \otimes e^{\mathfrak{i} P t} \otimes I,
\end{aligned}
\end{equation}
which is also straightforward to prove for any number of $I$ operators, or for any number of $I$ operators between two, or more, Pauli operators. \textcolor{black}{This describes the fact that a Hamiltonian acting nontrivially on some of the subsystems induces the unitary evolution of exclusively these subsystems, leaving the rest of the subsystems unchanged.}  Then, for $k<j$ it holds that the $k,j$-th Hamiltonian term is given as
\begin{equation} \label{Ham1} 
\begin{aligned}
 H_{kj} =&\,\,   Q_{kj} I ^ {\otimes(k-1)} \otimes e^{-\mathfrak{i}\sum_{i = 0}^{b-1}Z_{(k-1)b+i}\theta_i} \otimes I ^ {\otimes(j-k)} \\
 &\,\,\otimes e^{\mathfrak{i}\sum_{i = 0}^{b-1}Z_{(j-1)b+i}\theta_i} \otimes I ^ {\otimes(N_\mathrm{T}-j)} \\
 =&\,\, Q_{kj} I ^ {\otimes(k-1)} \bigotimes_{i=0}^{b-1} R_{Z_{(k-1)b+i}}(2\theta_i) \otimes I ^ {\otimes(j-k)} \\
 &\,\,\bigotimes_{i=0}^{b-1} R_{Z_{(j-1)b+i}}(-2\theta_i) \otimes I ^ {\otimes(N_\mathrm{T}-j)},
\end{aligned}
\end{equation}
while for the case $k>j$, the $k,j$-th term of the Hamiltonian is given as
\begin{equation} \label{Ham2}
\begin{aligned}
 \hat{H}_{kj} =&\,\,   Q_{kj} I ^ {\otimes(j-1)} \otimes e^{-\mathfrak{i}\sum_{i = 0}^{b-1}Z_{(j-1)b+i}\theta_i} \otimes I ^ {\otimes(k-j)} \\
 &\,\,\otimes e^{\mathfrak{i}\sum_{i = 0}^{b-1}Z_{(k-1)b+i}\theta_i} \otimes I ^ {\otimes(N_\mathrm{T}-k)} \\
 =&\,\, Q_{kj} I ^ {\otimes(j-1)} \bigotimes_{i=0}^{b-1} R_{Z_{(j-1)b+i}}(2\theta_i) \otimes I ^ {\otimes(k-j)} \\
 &\,\,\bigotimes_{i=0}^{b-1} R_{Z_{(k-1)b+i}}(-2\theta_i) \otimes I ^ {\otimes(N_\mathrm{T}-k)}.
\end{aligned}
\end{equation}
\textcolor{black}{To construct the quantum circuit corresponding to this cost Hamiltonian, first, we have to derive a relation for \(\bigotimes_{i=0}^{b-1} R_{Z_{(k-1)b+i}}(-2\theta_i)\) expressed purely in terms of Pauli operators. This is necessary to exrtact the unitary evolution of the Hamiltonian.} The tensor product of two rotation gates is already provided in \eqref{eq:rtensorr}. To generalize this to \(b\) rotation gates, we explicitly compute the tensor product for three rotation gates, illustrating the pattern that emerges for any \(b\). When three rotation gates are involved, their tensor product is given by
\begin{equation} \label{rrrrtensor}
\begin{aligned}
R_{Z_1}(\theta_1)\otimes &R_{Z_2}(\theta_2) \otimes R_{Z_3}(\theta_3) = \\
     &\cos\left(\frac{\theta_1}{2}\right) \cos\left(\frac{\theta_2}{2}\right) \cos\left(\frac{\theta_3}{2}\right) I \otimes I \otimes I \\
 - \,\mathfrak{i}\, &\sin\left(\frac{\theta_1}{2}\right) \cos\left(\frac{\theta_2}{2}\right) \cos\left(\frac{\theta_3}{2}\right) Z \otimes I \otimes I \\
 - \,\mathfrak{i} \,&\cos\left(\frac{\theta_1}{2}\right) \sin\left(\frac{\theta_2}{2}\right) \cos\left(\frac{\theta_3}{2}\right) I \otimes Z \otimes I \\
 - \,&\sin\left(\frac{\theta_1}{2}\right) \sin\left(\frac{\theta_2}{2}\right) \cos\left(\frac{\theta_3}{2}\right) Z \otimes Z \otimes I \\
 +\, \mathfrak{i} \,&\sin\left(\frac{\theta_1}{2}\right) \sin\left(\frac{\theta_2}{2}\right) \sin\left(\frac{\theta_3}{2}\right) Z \otimes Z \otimes Z \\
 - \,&\cos\left(\frac{\theta_1}{2}\right) \sin\left(\frac{\theta_2}{2}\right) \sin\left(\frac{\theta_3}{2}\right) I \otimes Z \otimes Z \\
 - \,&\sin\left(\frac{\theta_1}{2}\right) \cos\left(\frac{\theta_2}{2}\right) \sin\left(\frac{\theta_3}{2}\right) Z \otimes I \otimes Z \\
 - \,\mathfrak{i} \,&\cos\left(\frac{\theta_1}{2}\right) \cos\left(\frac{\theta_2}{2}\right) \sin\left(\frac{\theta_3}{2}\right) I \otimes I \otimes Z.
\end{aligned}
\end{equation}
Observing \eqref{eq:rtensorr} and \eqref{rrrrtensor}, we can induce the following
\begin{equation}
\begin{aligned}
&\bigotimes_{i=1}^n R_{Z_{i}}(\theta_i) = \\
& \sum_{\boldsymbol{\alpha}\in\!\{0,1\}^{ n}}\!\!\!\!\!
(-\mathfrak{i})^{|\boldsymbol{\alpha}|}
\!\prod_{j=1}^{n}
\cos\bigl(\tfrac{\theta_j}{2}\bigr)^{1-\alpha_j}\!\sin\bigl(\tfrac{\theta_j}{2}\bigr)^{\alpha_j}\!\left(Z^{\alpha_1}\!\otimes\!\cdots\!\otimes\!Z^{\alpha_n}\right),
\end{aligned}
\end{equation}
where $\boldsymbol{\alpha} = (\alpha_1,\dots,\alpha_n)$, $|\boldsymbol{\alpha}| = \alpha_1+\dots+\alpha_n$ is the Hamming weight of $\boldsymbol{\alpha}$, while the summation spans all possible binary strings of length $n$. Also, $Z^0 = I$ and $Z^1 = Z$. \textcolor{black}{This expansion makes explicit that the global, $n$-qubit $Z$-rotation admits a decomposition as a weighted sum of all possible Pauli-$Z$ strings of length $n$.} It is also noted that $\bigotimes_{i=1}^n R_{Z_{i}}(-\theta_i)$ is given the same way, with the exception of $\mathfrak{i}^{|\boldsymbol{\alpha}|}$ being used, instead of $(-\mathfrak{i})^{|\boldsymbol{\alpha}|}$. As such, the final expression for the $k,j$-th term of the Hamiltonian, for $k<j$, is the following
\begin{equation} \label{Ham1}
\begin{aligned}
 H_{kj} &= Q_{kj}\!\!\!\!\!\!\sum_{\boldsymbol{\alpha}\in\!\{0,1\}^{ 2b}}\!\!\!\!\!
\mathfrak{i}^{\left(2|\boldsymbol{\alpha}_{[:b]}|+|\boldsymbol{\alpha}_{[b:]}|\right)} \!\prod_{j=1}^{2b}
\cos\bigl(\theta_j\bigr)^{1-\alpha_j}\!\sin\bigl(\theta_j\bigr)^{\alpha_j} \\
& \times I ^ {\otimes(k-1)} \!\otimes Z^{\alpha_1}\!\otimes\!\cdots\!\otimes\!Z^{\alpha_b} \otimes I ^ {\otimes(j-k)} \\
&\otimes Z^{\alpha_{b+1}}\!\otimes\!\cdots\!\otimes\!Z^{\alpha_{2b}} \otimes I ^ {\otimes(N_\mathrm{T}-j)},
\end{aligned}
\end{equation}
where $|\boldsymbol{\alpha}_{[:b]}|$ is the Hamming weight of the first half of the string, and $|\boldsymbol{\alpha}_{[b:]}|$ the Hamming weight of the second half,
while the final expression of $\hat{H}_{kj}$, for $k>j$, is given in a similar fashion and it is omitted. \textcolor{black}{We note that this expression occurred by considering the tensor product of both $\bigotimes_{i=1}^n R_{Z_{i}}(\theta_i)$  and $\bigotimes_{i=1}^n R_{Z_{i}}(-\theta_i)$, while $|\boldsymbol{\alpha}_{[:b]}|$ is the number of $Z$ gates associated with the Pauli strings of $\bigotimes_{i=1}^n R_{Z_{i}}(\theta_i)$ and $|\boldsymbol{\alpha}_{[b:]}|$ is the number of $Z$ gates associated with the Pauli strings of $\bigotimes_{i=1}^n R_{Z_{i}}(-\theta_i)$.}
Therefore, the cost Hamiltonian for the general problem is given as
\begin{equation} \label{Ham_final}
    H_C = \sum_{k = 1}^{N_\mathrm{T}}\sum_{\substack{j=1 \\ j \neq k}}^{N_\mathrm{T}} \mathbb{I}(k<j){H}_{kj} + \mathbb{I}(k>j)\hat{H}_{kj},
\end{equation}
where \(\mathbb{I}(\cdot)\) is an indicator function that equals one if its argument is true, and zero otherwise. 

We note that all the terms in \(H_C\) commute. This is straightforward to prove because \(H_C\) is a sum of terms of the form $I^{\otimes (k-1)} \otimes Z^{\alpha_1} \otimes \cdots \otimes Z^{\alpha_b} \otimes I^{\otimes (j-k)} \otimes Z^{\alpha_{b+1}} \otimes \cdots \otimes Z^{\alpha_{2b}} \otimes I^{\otimes (N_\mathrm{T} - j)}$, where each term is diagonal w.r.t. the computational basis; since both \(I\) and \(Z\) are diagonal, so is any tensor product of them. 
Then, we can obtain the quantum operator $U_C$. Abstractly, after all additions in \eqref{Ham_final}, the cost Hamiltonian is given as
\begin{equation}
\begin{aligned}
    H_C =& \sum_{k = 1}^{N_\mathrm{T}}\sum_{\substack{j=1 \\ j \neq k}}^{N_\mathrm{T}}\sum_{\boldsymbol{\alpha}\in\!\{0,1\}^{ 2b}} Q_{\boldsymbol{\alpha}}H_{k,j}^{\boldsymbol{\alpha}}, \\ 
    H_{k,j}^{\boldsymbol{\alpha}} =& \,I ^ {\otimes(\max(k,j)-1)} \!\otimes Z^{\alpha_1}\!\otimes\!\cdots\!\otimes\!Z^{\alpha_b} \otimes I ^ {\otimes(|j-k|)} \\
&\otimes Z^{\alpha_{b+1}}\!\otimes\!\cdots\!\otimes\!Z^{\alpha_{2b}} \otimes I ^ {\otimes(N_\mathrm{T}-\min(k,j))},
\end{aligned}
\end{equation}
where $Q_{\boldsymbol{\alpha}}$ is a real number occurred after all additions, while $H_{k,j}^{\boldsymbol{\alpha}}$ are the Pauli tensor products involved. Then, the following holds
\begin{equation} \label{eq:Uc} 
\begin{aligned}
    U_C =\,\, & \prod_{k = 1}^{N_\mathrm{T}}\prod_{\substack{j=1 \\ j \neq k}}^{N_\mathrm{T}}\prod_{\boldsymbol{\alpha}\in\!\{0,1\}^{ 2b}} e^{-\mathfrak{i}\gamma Q_{\boldsymbol{\alpha}}H_{k,j}^{\boldsymbol{\alpha}}},
\end{aligned}
\end{equation}
where this exact factorization is allowed because the exponential of a sum of commuting operators is equal to the product of their exponential, thus the Trotter–Suzuki decomposition is not needed. Although it is feasible to explicitly compute the value of \(U_C\), doing so is rather tedious and unnecessary for implementing the QAOA algorithm. Instead, we note that \eqref{eq:Uc} can be easily implemented in specialized programming languages, such as Qiskit \cite{javadi2024quantum}. Nonetheless, we note that the explicit formulation of the cost Hamiltonian is required in order to calculate $U_C$. Moreover, the analysis w.r.t. the optimization vector $\boldsymbol{g}$ is the same by using $N_\mathrm{R}$ instead of $N_\mathrm{T}$ and $A$ as it was defined in section IV. The algorithmic procedure for maximizing the SNR is summarized in Algorithm 2.

\subsection{The WS-QAOA approach}
Based on whether we optimize w.r.t. to $\boldsymbol{f}$ or $\boldsymbol{g}$, $Q$ is defined accordingly. Then, the relaxed beamforming optimization problem is given as follows
\begin{equation}\label{eq:opt_problem}
\begin{aligned}
\max_{\mathbf{x}} \quad & \sum_{k=1}^{N_\mathrm{T}}\sum_{\substack{j=1 \\ j \neq k}}^{N_\mathrm{T}} Q_{kj}\, e^{\mathfrak{i}\left( -\sum_{i=1}^{b} x_i^k\,\theta_i + \sum_{i=1}^{b} x_i^j\,\theta_i \right)} \\
\text{s.t.} \quad & x_i^k \in [0,1], \,\forall\, k = 1,\dots,N_\mathrm{T},\, i = 1,\dots,b.
\end{aligned}
\end{equation}
The objective function in this problem is real-valued, enabling the application of any gradient-based optimization algorithm to determine its optimal solution, \(\mathbf{x}^*\), or equivalently \(\boldsymbol{c}^*\), as it was defined in Section II. \textcolor{black}{ The optimization process is performed with respect to \(\boldsymbol{f}\) and \(\boldsymbol{g}\) in a successive manner until convergence is achieved. Once the relaxed problem is solved, the best solutions \(\mathbf{x}_f^*\) and \(\mathbf{x}_g^*\) are obtained. These solutions, along with the relations defined in equations (10) and (11), are used to compute the initial quantum state and the mixer Hamiltonian for the WS-QAOA approach. We note that the cost Hamiltonian is unchanged for both schemes. As a consequence, lines 1 and 6–7 in Algorithm 2 are the only parts that differ between the WS-QAOA based and the QAOA based scheme. Beyond these differences, the rest of the algorithmic procedure is identical in both approaches.} \textcolor{black}{We note that the notation $\boldsymbol{f}^*,\boldsymbol{g}^*$ again denotes the best solution found, which may not necessarily be optimal. The performance of Alg. 2 will also be numerically evaluated.}

\begin{algorithm}[t]
\caption{$b$-bit QAOA-based beamforming optimization}
\textbf{Input:} Matrix $H \in \mathbb{C}^{N_\mathrm{T} \times N_\mathrm{R}}$, initial vector $\mathbf{g}^0\in\mathbb{S}^{N_\mathrm{R}}$, maximum number of iterations $K$, QAOA layers $p$.
\begin{algorithmic}[1]
    \State \parbox[t]{\dimexpr\linewidth-\algorithmicindent}{
        \textbf{For WS-QAOA}: Solve \eqref{eq:opt_problem} to obtain $\mathbf{x}_f^*,\ \mathbf{x}_g^*$, then set 
        $\mathbf{g}^{0}\gets \mbox{round}(\mathbf{x}_g^*)$ or use a random feasible value.
    } \vspace{.1cm}
    \State \parbox[t]{\dimexpr\linewidth-\algorithmicindent}{
        Set $\theta_j=2^j\frac{2\pi}{2^b}$ for $j=0,1,\dots,b-1$, and define $\boldsymbol{f}$ and $\boldsymbol{g}$ as in (28)
    } \vspace{.1cm}
    \For{$k=1,\dots,K$}
        \State \parbox[t]{\dimexpr\linewidth-\algorithmicindent}{
            \textbf{Solving w.r.t. $\boldsymbol{f}$:} If $k=1$, set $\mathbf{g}\gets \mathbf{g}^0$; otherwise, set $\mathbf{g}\gets \mathbf{g}^*$.
        } \vspace{.05cm}
        \State \parbox[t]{\dimexpr\linewidth-\algorithmicindent}{
            Set $ A=\mathbf{g}^{*\dagger}H$, $Q= A^\dagger  A$, and compute $H_C$ per (48).
        }
        \State \parbox[t]{\dimexpr\linewidth-\algorithmicindent}{
            Define the mixer Hamiltonian:
            \begin{itemize}[leftmargin=2em, label={}]
                \item \textbf{For QAOA}: $H_M=\sum_{j=1}^{bN_\mathrm{T}}X_j$.
                \item \textbf{For WS-QAOA}: use (11) with $\mathbf{x}_f^*$ and initialize $y_i = 2 \arcsin(\sqrt{\mathbf{x}_{f_i}^*})$.
            \end{itemize}
        }
        \State \parbox[t]{\dimexpr\linewidth-\algorithmicindent}{
            Initialize $\ket{\psi_0}$ in $\ket{+}^{\otimes bN_\mathrm{T}}$ or in $\bigotimes_{i=1}^{bN_\mathrm{T}}R_{Y_i}(y_i)\lvert0\rangle$.
        }
        \For{$\ell=1,\dots,p$}
            \State \parbox[t]{\dimexpr\linewidth-\algorithmicindent}{Apply $U_C(\gamma_\ell)=\exp\bigl(-\mathfrak{i}\,\gamma_\ell\,H_C\bigr)$.}
            \State \parbox[t]{\dimexpr\linewidth-\algorithmicindent}{Apply $U_M(\beta_\ell)=\exp\bigl(-\mathfrak{i}\,\beta_\ell\,H_M\bigr)$.}
        \EndFor
        \State \parbox[t]{\dimexpr\linewidth-\algorithmicindent}{
            Maximize 
            \[
            F(\boldsymbol{\gamma},\boldsymbol{\beta}) = \langle \psi(\boldsymbol{\gamma},\boldsymbol{\beta})|H_C|\psi(\boldsymbol{\gamma},\boldsymbol{\beta})\rangle
            \]
            via a classical optimizer. \vspace{.1cm}
        }
        \State \parbox[t]{\dimexpr\linewidth-\algorithmicindent}{
            After convergence, measure to obtain bitstrings $\mathbf{x}^*$; select the best candidate based on SNR.
        }
        \\
        \State \parbox[t]{\dimexpr\linewidth-\algorithmicindent}{
            \textbf{Solving w.r.t. $\boldsymbol{g}$:} Set $\boldsymbol{f}\gets \boldsymbol{f}^*$, $A=H\boldsymbol{f}$ and $Q=A^\dagger A$, then follow a similar approach to obtain $\boldsymbol{g}^*$. \vspace{.1cm}
        }
    \EndFor
    \State \parbox[t]{\dimexpr\linewidth-\algorithmicindent}{
        \textbf{Output:} $(\boldsymbol{f}^*,\boldsymbol{g}^*)=\arg\max_{\boldsymbol{f},\boldsymbol{g}}\rho(\boldsymbol{f},\boldsymbol{g})$.
    }
\end{algorithmic}
\end{algorithm}

\section{Complexity analysis}
In this paper three optimization approaches will be compared, the QAOA, the exhaustive search, and the quantized singular value decomposition (SVD). As such, the complexity of all schemes will be explained. To theoretically compare the complexity of all schemes we will measure the number of operations implemented in a classical and a quantum computer, where in the latter case this is approximated by the number of necessary quantum gates. The exhaustive search scheme evaluates the SNR expression for all the possible transmit and receive vectors and returns the solution with the maximum SNR. The exhaustive search approach requires $N_\mathrm{T}N_\mathrm{R}2^{\frac{b}{2}(N_\mathrm{T}+N_\mathrm{R})}$ computations, since it needs to reevaluate the product $\left| \mathbf{g}^{\dagger} H \mathbf{f} \right|^2$ a total of $2^{\frac{b}{2}(N_\mathrm{T}+N_\mathrm{R})}$ times. Therefore its complexity becomes exponential with the number of transmit/receive antennas, which is prohibited for large-scale MIMO topologies.
For a dense matrix \(H \in \mathbb{R}^{N_\mathrm{T} \times N_\mathrm{R}}\), the computational complexity of the full SVD using classical methods, e.g., the Golub–Reinsch algorithm, is typically \(\mathcal{O}(N_\mathrm{T} N_\mathrm{R}^2)\) when \(N_\mathrm{T} \geq N_\mathrm{R}\) (or \(\mathcal{O}(N_\mathrm{R} N_\mathrm{T}^2)\) if \(N_\mathrm{T} < N_\mathrm{R}\)) \cite{svd}; however, if only the top \(k\) singular values and corresponding singular vectors are needed (with \(k \leq \min\{N_\mathrm{T},N_\mathrm{R}\}\)), iterative methods such as Lanczos bidiagonalization or randomized SVD can be utilized, often reducing the complexity to approximately \(\mathcal{O}(kN_\mathrm{T} N_\mathrm{R})\). In addition, we consider the quantization procedure to have a worst-case required time complexity of roughly \( \mathcal{O}(b(N_\mathrm{T}+N_\mathrm{R})) \), which can be omitted. 

To evaluate the complexity of QAOA on a fully connected-hardware, we must consider both the quantum circuit and the classical optimizer \cite{guerreschi2019qaoa}. The quantum circuit's complexity is typically measured in terms of the number of gates, which is a common metric for assessing the runtime of a quantum algorithm.  From (49) we observe that the operator corresponding to the cost Hamiltonian contains $2^bN_\mathrm{T}^2$ terms. Each such term is the exponential of $H_{k,j}^{\boldsymbol{\alpha}}$, which always contains the tensor product of $2b$ $Z$ gates with several $I$ gates. Generalizing (37) to the case of $2b$ $Z$ gates, while omitting the $I$ gates which is a direct result of (41), we can conclude that every  $H_{k,j}^{\boldsymbol{\alpha}}$ term requires $2b$ CNOT gates plus one $R_Z$ gate. As such the circuit complexity for the quantum circuit modeling the cost Hamiltonian of AO subproblems is  $\mathcal{O}(b2^b(N_\mathrm{T}^2+N_\mathrm{R}^2))$ gates. With the total number of qubits given by \(bN_\mathrm{T}\), the typical state initialization is achieved via Hadamard gates and the mixer Hamiltonian, which is implemented by single-qubit rotations such as \(R_Z\), where each contribute \(\mathcal{O}(bN_\mathrm{T})\) gates. In most practical regimes, \(b\,2^b\,N_\mathrm{T}\,N_\mathrm{R} \gg bN_\mathrm{T}\), so the cost Hamiltonian dominates the per-layer cost. In the WS-QAOA framework, additional rotation gates required for state preparation are similarly counted as single-qubit gates, preserving the overall scaling. Thus, for a QAOA circuit with \(p\) layers, the  gate count scales as $ \mathcal{O}\left(pb2^b(N_\mathrm{T}^2+N_\mathrm{R}^2)\right)$ \cite{guerreschi2019qaoa}.

The COBYLA optimizer builds a linear approximation of the objective function using interpolation points, where for \(2p\) parameters the full set of points is given by $
m =(p+1)(2p+1) = \mathcal{O}(p^2)$, implying that \(m = \mathcal{O}(p^2)\) \cite{powell2007view}. According to the analysis, each iteration of COBYLA requires \(\mathcal{O}(m^2)\) function evaluations, leading to a per-iteration complexity of \(\mathcal{O}(p^4)\) \cite{powell2007view}.  Therefore, the overall complexity of the QAOA and WS-QAOA algorithm is $\mathcal{O}\left(Kp^5b2^b(N_\mathrm{T}^2+N_\mathrm{R}^2)\right)$, where $K$ is the number of AO iterations. Please note that in some extensions and practical implementations, such as those seen in algorithms like the NEWUOA, one can reduce the number of interpolation points used to build the local model. For instance, instead of using the full set of \(\mathcal{O}(p^2)\) points, a reduced set of \(2p+1\) points may be employed to construct a quadratic approximation. This reduction diminishes the computational overhead per iteration, since the cost of updating the model becomes proportional to the square of the number of points, which is now \(\mathcal{O}(p^2)\) rather than \(\mathcal{O}(p^4)\).

\section{Numerical Results}
In this section, numerical results are presented to verify the effectiveness of QAOA on addressing the $b$-bit digital beamforming optimization problem.  \textcolor{black}{We note that to verify the performance of the QAOA-based approach w.r.t. the optimal solution, we consider topologies for which the exhaustive-search is feasible. Large-scale simulations are left as an extension of this work, since advanced QAOA approaches and error-correction mechanism, which will build on the current analysis, will need to be introduced.} The number of antennas and the  number of quantization bits $b$ are given in each figure.  Moreover,  we assume normalized Rayleigh block fading channels, i.e., $h_{i,j} \sim \mathcal{CN}(0,1)$ and $P = 1$, $\sigma^2 = 1$. We note that the selection of $P,\sigma^2$ do not affect the optimal solution of problem (13). For the QAOA procedure we set $K = 5$ and $p = 3$. All results were averaged on 100 Monte Carlo iterations, so for instance, a $5\times5$ MIMO simulation involves 2500 different channels. Also, all channels were created prior to the simulation, so all benchmarks were tested on the exact channel conditions for all Monte Carlo iterations. \textcolor{black}{Furthermore, the simulations in this work were conducted using IBM's FakeKyoto backend, provided through Qiskit's fake provider module \cite{javadi2024quantum}. This backend mimics the behavior of IBM Quantum systems based on system snapshots, which includes details such as the coupling map, basis gates, and qubit properties, such as error rates. The noise model used in our simulations is derived from these snapshots, ensuring a realistic representation of hardware imperfections.} All considered schemes are outlined below:
\begin{itemize}
    \item Exhaustive search: This scheme evaluates the SNR expression for all the possible transmit and receive vectors and returns the solution with the maximum SNR.
    \item \textcolor{black}{SA: SA is a classical optimization algorithm inspired by the annealing process in metallurgy. It probabilistically explores the  solution space by gradually decreasing a temperature parameter, allowing occasional acceptance of worse solutions to escape local minima.}
    \item Quantized SVD: This scheme leverages the full SVD decomposition of the channel matrix to identify the optimal pre- and post-coding vectors and gives as its output the quantized optimal vectors. 
    \item Alg. 1: This scheme gives the solution obtained from running Alg. 1 for various different initial points. It is applicable only for $b = 2$.
    \item Alg. 2 (no warm-start): This is the solution given from the designed QAOA procedure.
    \item Alg. 2 (warm-start): This is the solution given from the designed WS-QAOA procedure. 
    \item \textcolor{black}{CO-based: A convex optimization approach is evaluated, utilizing AO and quantization, based on \cite{krikidis1, 7397861}.}
\end{itemize}

\begin{figure}
    \centering
    \begin{tikzpicture}
        \begin{axis}[
            scaled ticks=false, 
            tick label style={/pgf/number format/fixed},
            width=\linewidth,
            xlabel = {$N_\mathrm{T}\times N_\mathrm{R}$},
            ylabel = {Average $\rho(\mathbf{f}^*,\mathbf{g}^*)$},
            ymin = 0,
            ymax = 120,
            xmin = 2,
            xmax = 10,
            xtick={2,4,6,8, 10},
            xticklabels = {$2\times2$,$4\times4$,$6\times6$,$8\times8$,$10\times10$},
            ytick = {20,40,60,80,100,120},
            yticklabels = {10,20,30,40,50,60},
            grid = major,
            legend entries = {\textcolor{black}{Infinite resolution}, {Exhaustive search, \textcolor{black}{SA}}, Alg. 2 (warm-start), Alg. 1 (3 initial points), Alg. 1 (1 initial point), Quantized SVD, \textcolor{black}{CO-based}},
            legend cell align = {left},
            legend style={at={(0,1)},anchor=north west, font=\scriptsize},
        ]
            \addplot[
                black,
                mark = none,
                line width = 1pt,
                style = solid,
            ]
            table {figs/inf_resol_transformed.dat};
            \addplot[
                black,
                mark = o,
                line width = 1pt,
                style = solid,
            ]
            table {SNR_vs_NxN_Brute_Force.dat};

            \addplot[
                red,
                mark = star,
                line width = 1pt,
                style = solid,
            ]
            table {SNR_vs_NxN_Alg_2.dat};
            
            \addplot[
                black,
                mark = triangle*,
                line width = 1pt,
                style = solid,
            ]
            table {SNR_vs_NxN_QAOA.dat};
            \addplot[
                black,
                mark = diamond*,
                line width = 1pt,
                style = solid,
            ]
            table {SNR_vs_NxN_Alg_1_one_init.dat};

            \addplot[
                black,
                mark = square,
                line width = 1pt,
                style = solid,
            ]
            table {Quantized_SVD.dat};
            \addplot[
                black,
                mark = *,
                line width = 1pt,
                style = solid,
            ]
            table {figs/copt_2.dat};
        \end{axis}
    \end{tikzpicture}
    \caption{Average $\rho(\mathbf{f}^*, \mathbf{g}^*)$ vs $N_\mathrm{T}\times N_\mathrm{R}$, $b=2$.}
    \label{fig:6}
\end{figure}
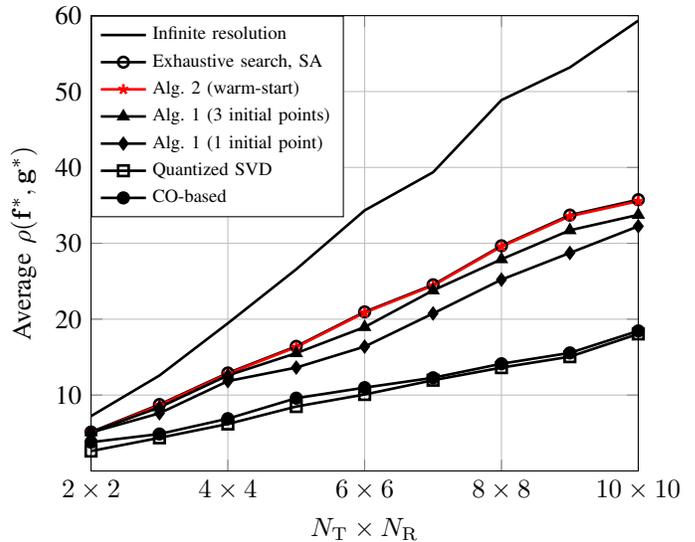

In Fig. 2, the scenario for \( b = 2 \) is presented for various number of antennas. It is evident that the Quantized SVD scheme underperforms compared to all other methods, primarily due to the low number of quantization bits, which restricts its precision.  Algorithm 1, on the other hand, achieves results that are near-optimal as the number of initial points is increased. However, this improvement comes at the cost of higher computational complexity, as adding more initial points directly increases the overall runtime of the algorithm. 
To better understand the behavior of Algorithm 1, we refer to the QAOA landscape shown in Fig. 5, defined by \( F(\boldsymbol{\gamma}, \boldsymbol{\beta}) = \langle \psi(\boldsymbol{\gamma}, \boldsymbol{\beta}) | H_C | \psi(\boldsymbol{\gamma}, \boldsymbol{\beta}) \rangle \). The landscape is characterized by its rugged and highly irregular structure, with numerous poor-quality local maxima. These features often trap the algorithm in suboptimal regions, preventing it from reaching the global optimum. 
In contrast, Algorithm 2 consistently performs at an optimal level. This superior performance can be attributed to its smoother and less fragmented landscape, which reduces the presence of suboptimal local maxima and facilitates convergence to the global maximum. At first glance, the difference in the landscapes of the two problems may seem unexpected, as for $b=2$ the two algorithms are theoretically expected to yield the same result. However, this is not the case. The discrepancy arises from the differences in how the two algorithms approach the problem. In Algorithm 1, the QAOA is treated as a black-box optimizer, with algebraic manipulations performed to separate the real and imaginary components of the optimization variables. In contrast, Algorithm 2 directly incorporates the problem's structure into the Hamiltonian. \textcolor{black}{ Moreover, we note that the CO-based benchmark struggles to find a high-quality solution for this problem. This limitation arises primarily because, in each round of AO, the algorithm first identifies a continuous solution, which is subsequently quantized into a feasible discrete solution. Such quantization introduces significant optimization errors, resulting in a considerable deviation from optimality.}


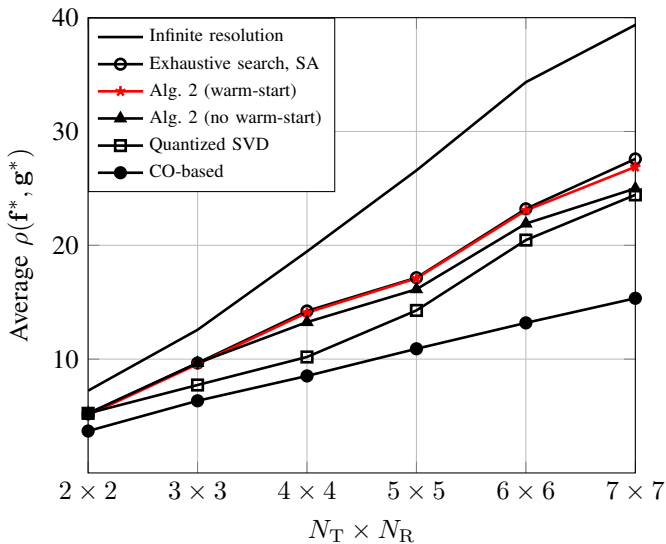
\begin{figure}
    \centering
    \begin{tikzpicture}
        \begin{axis}[
            scaled ticks=false, 
            tick label style={/pgf/number format/fixed},
            width=\linewidth,
            xlabel = {$N_\mathrm{T}\times N_\mathrm{R}$},
            ylabel = {Average $\rho(\mathbf{f}^*,\mathbf{g}^*)$},
            ymin = 0,
            ymax = 40,
            xmin = 2,
            xmax = 7,
            xtick={2,3,4,5,6,7},
            xticklabels={$2\times2$,$3\times3$,$4\times4$,$5\times5$,$6\times6$,$7\times7$},
            ytick = {10,20,30,40},
            yticklabels = {10,20,30,40},
            grid = major,
            legend entries = {\textcolor{black}{Infinite resolution}, {Exhaustive search, \textcolor{black}{SA}}, Alg. 2 (warm-start), Alg. 2 (no warm-start), Quantized SVD, \textcolor{black}{CO-based}},
            legend cell align = {left},
            legend style={at={(0,1)},anchor=north west, font=\scriptsize},
        ]
            \addplot[
                black,
                mark = none,
                line width = 1pt,
                style = solid,
            ]
            table {figs/inf_resol.dat};
            \addplot[
                black,
                mark = o,
                line width = 1pt,
                style = solid,
            ]
            table {figs/b_3/brute_force_3.dat};

            \addplot[
                red,
                mark = star,
                line width = 1pt,
                style = solid,
            ]
            table {figs/b_3/qaoa_with_warm_start_3.dat};
            
            \addplot[
                black,
                mark = triangle*,
                line width = 1pt,
                style = solid,
            ]
            table {figs/b_3/qaoa_no_warm_start_3.dat};
            \addplot[
                black,
                mark = square,
                line width = 1pt,
                style = solid,
            ]
            table {figs/b_3/quantized_svd_3.dat};
            \addplot[
                black,
                mark = *,
                line width = 1pt,
                style = solid,
            ]
            table {figs/copt_3.dat};
        \end{axis}
    \end{tikzpicture}
    \caption{Average $\rho(\mathbf{f}^*, \mathbf{g}^*)$ vs $N_\mathrm{T}\times N_\mathrm{R}$, $b=3$.}
    \label{fig:6}
\end{figure}

In Fig. 3, we illustrate the performance for \( b = 3 \). This time, we note that the quantized SVD performs relatively better due to the increased value of \( b \), which enhances its representation accuracy. Nonetheless, the SVD scheme still performs considerably worse than the optimal solution. Algorithm 2 without a warm-start performs reasonably; however, its performance deteriorates as the number of antennas increases. This is because the search space of the QAOA algorithm expands, making it difficult for the algorithm to find the optimal solution with the given QAOA depth \( p \). This behavior is also well explained by the QAOA landscape of the algorithm. Specifically, while the landscape contains only two high-quality local maxima, the remainder is a barren plateau that prevents the classical optimizer from progressing toward the optimum. In fact, barren plateaus are a common issue in QAOA-based implementations. In contrast, Algorithm 2 with a warm-start consistently finds the optimal solution. This improvement can also be explained by the landscape. Following the procedure of Section II.B, the implemented algorithm yields a landscape similar to the one shown in Fig. 5(c), which is smooth, with few local maxima, most of which are high quality, thus the algorithm being trapped in one of these causes little, if any, performance loss. Moreover, the absence of barren plateaus favors classical optimizers, as reflected in its enhanced performance. \textcolor{black}{Furthermore, SA successfully reaches the optimal solution for both $b = 2$ and $b = 3$. However, it required several runs to tune the cooling hyperparameter and its initial temperature, which involved significant computational effort.}

\begin{figure}[t]
    \centering
    \begin{tikzpicture}
        \begin{axis}[
            scaled ticks=false, 
            tick label style={/pgf/number format/fixed},
            width=\linewidth,
            xlabel = {Quantization level, $b$},
            ylabel = {Average $\rho(\mathbf{f}^*,\mathbf{g}^*)$},
            ymin = 2,
            ymax = 12,
            xmin = 1,
            xmax = 5,
            xtick={1,2,3,4,5},
            xticklabels={1,2,3,4,5},
            ytick = {2,4,6,8,10,12},
            yticklabels = {2,4,6,8,10,12},
            grid = major,
            legend entries = {\textcolor{black}{Infinite resolution}, {Exhaustive search, \textcolor{black}{SA}}, Alg. 2 (warm-start), Alg. 2 (no warm-start), Quantized SVD, \textcolor{black}{CO-based}},
            legend cell align = {left},
            legend style={at={(1,0)},anchor=south east, font=\scriptsize},
        ]
        \addplot[
                black,
                mark = none,
                line width = 1pt,
                style = solid,
            ]
            table {figs/hor_line.dat};
            
            \addplot[
                black,
                mark = o,
                line width = 1pt,
                style = solid,
            ]
            table {figs/varying_beta/b3_brute_force.dat};

            \addplot[
                red,
                mark = star,
                line width = 1pt,
                style = solid,
            ]
            table {figs/varying_beta/b3_warm_start.dat};
            
            \addplot[
                black,
                mark = triangle*,
                line width = 1pt,
                style = solid,
            ]
            table {figs/varying_beta/be_no_warm_start.dat};
            \addplot[
                black,
                mark = square,
                line width = 1pt,
                style = solid,
            ]
            table {figs/varying_beta/b3_qsvd.dat};
            \addplot[
                black,
                mark = *,
                line width = 1pt,
                style = solid,
            ]
            table {figs/copt_b.dat};
        \end{axis}
    \end{tikzpicture}
    \caption{Average $\rho(\mathbf{f}^*, \mathbf{g}^*)$ vs $b$, $N_\mathrm{T}\times N_\mathrm{R}=3\times3$.}
    \label{fig:6}
\end{figure}

In Fig.~4, we evaluate the performance of all schemes in a \(3\times3\) MIMO scenario as the parameter \(b\) varies. Low-resolution digital beamforming, for \( b = 1 \) and \( b = 2 \), experiences approximately \( 50\% \) and \( 20\% \) performance degradation, respectively, highlighting the tradeoff between hardware complexity and performance. Nonetheless, \( \beta = 2 \) and $\beta = 3$ appear to strike a good balance between these factors. Furthermore, as expected, the Quantized SVD performs poorly for small values of \(b\). However, as \(b\) increases its performance approaches that of the other schemes, while the QAOA-based solution performs great in all scenarios. It should be noted that as \(b\) grows, the theoretical complexity of the QAOA algorithm can become higher than that of the full SVD approach. Approximately, the QAOA complexity exceeds that of the full SVD when
\begin{equation*} 
    b\,2^bp^3\,(N_\mathrm{T}^2 + N_\mathrm{R}^2) > N_\mathrm{T} N_\mathrm{R}^2.
\end{equation*}
Assuming \(N_\mathrm{T} = N_\mathrm{R}\) for simplicity, this condition implies that
\begin{equation*}
    \textcolor{black}{b > \frac{W_0(\frac{2N_\mathrm{T}}{p^3}\ln 2)}{\ln2 },}
\end{equation*}
where \(W_0(\cdot)\) denotes the principal branch of the Lambert \(W\) function. \textcolor{black}{As a consequence, when $b\le3$, our QAOA scheme has complexity on par with the quantized SVD.  Because both the CO-based and quantized SVD methods perform poorly at these low bit resolutions, QAOA offers a clear advantage in this regime.  In contrast, as $b$ increases towards full resolution, quantized SVD becomes the preferred approach.
}

\begin{figure}[htbp]              
  \centering
  \begin{subfigure}{\linewidth}   
    \centering
    \includegraphics[width=.8\columnwidth]{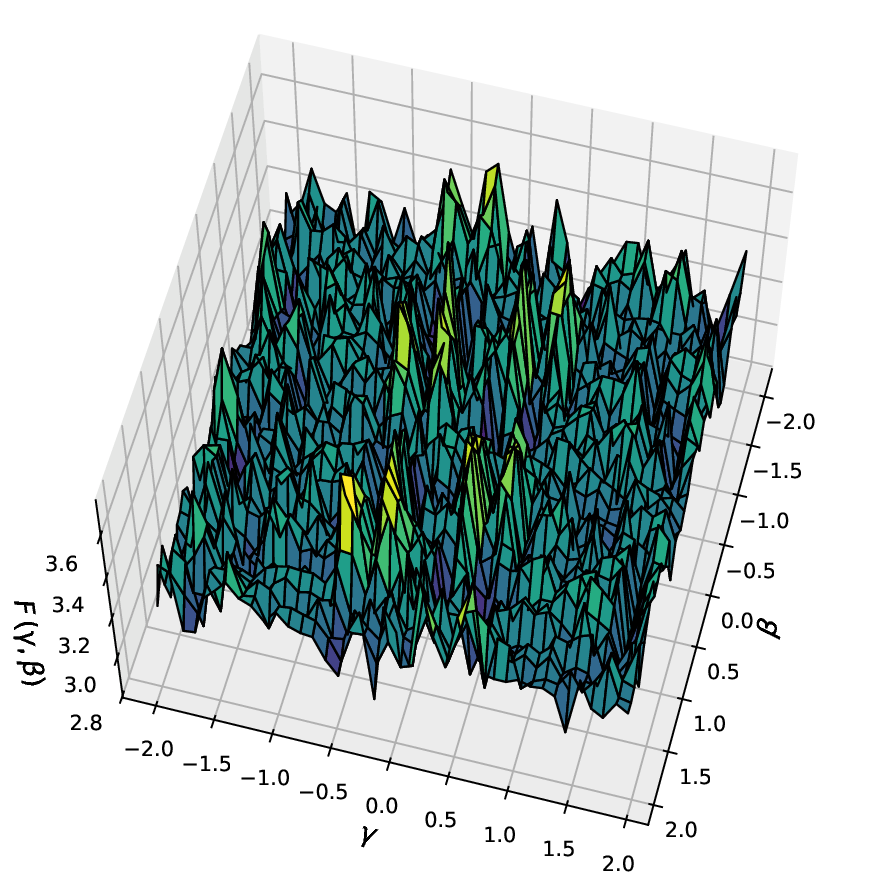}
    \caption{Algorithm 1.}
  \end{subfigure}
  \begin{subfigure}{\linewidth}
    \centering
    \includegraphics[width=.8\columnwidth]{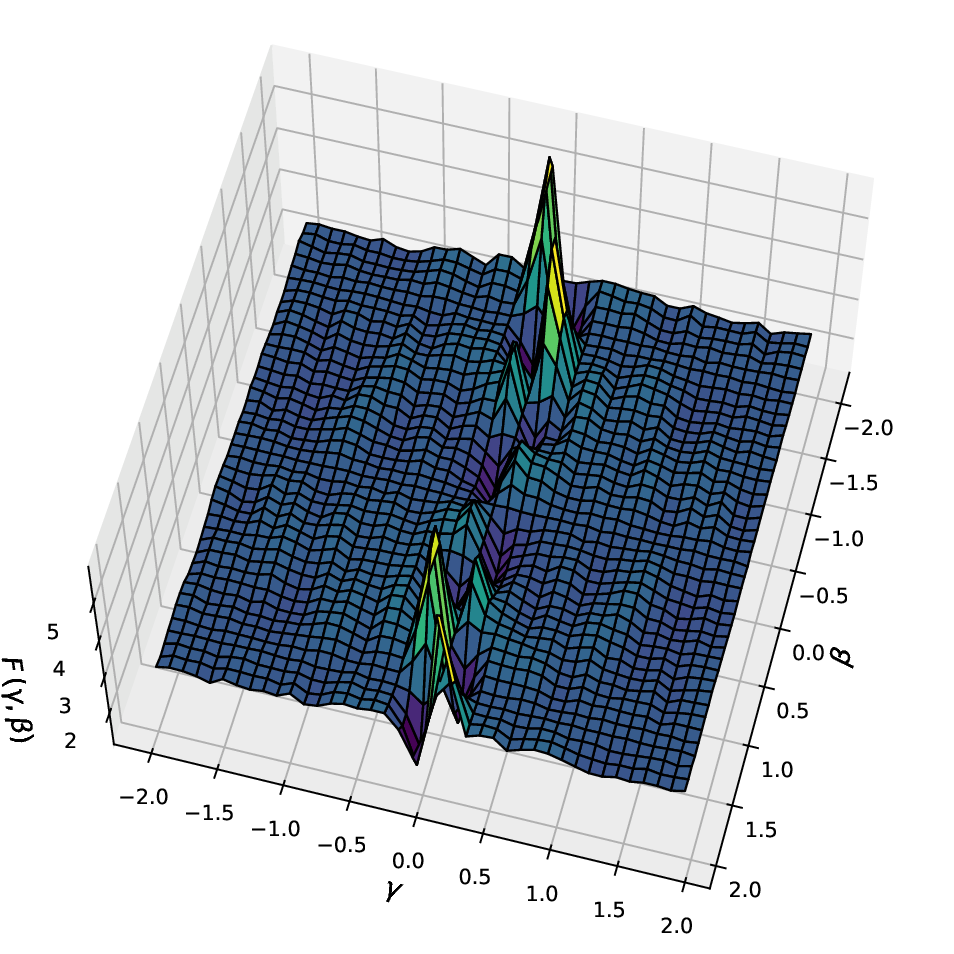}
    \caption{Algorithm 2, no warm-start.}
  \end{subfigure}
  \begin{subfigure}{\linewidth}
    \centering
    \includegraphics[width=.8\columnwidth]{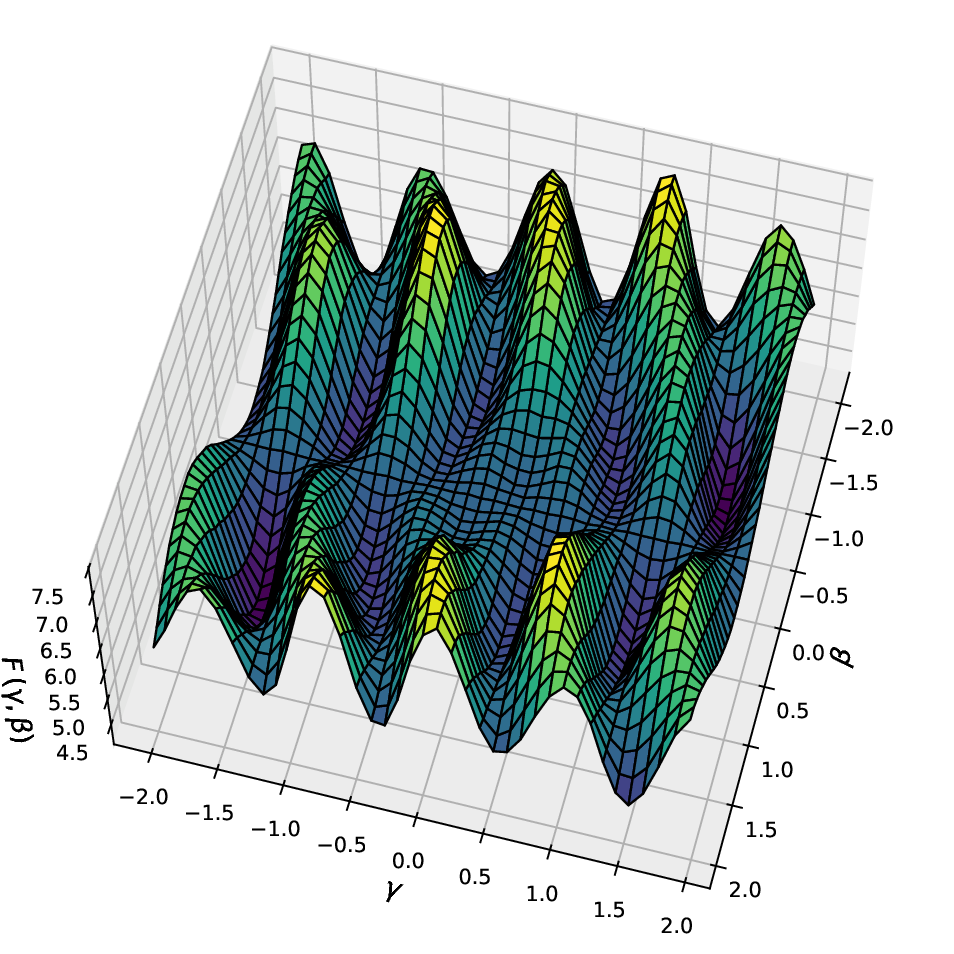}
    \caption{Algorithm 2, with warm-start.}
  \end{subfigure}
  \caption{Different QAOA landscapes for a \(5\times5\) MIMO scenario with \(b = 2\).}
  \label{fig:subfigures}
\end{figure}

Finally, Fig. 6 illustrates how the quasi-probability distribution of the QAOA evolves through the AO procedure. In the first iteration, the distribution is highly dispersed, reflecting a broad exploration of the solution space with no dominant outcomes. As the algorithm progresses to the second iteration with alternating updates on parameter subsets, the distribution begins to narrow, and certain outcomes, notably the red bitstring representing the optimal solution in this example, start to gain prominence, which is an early sign of convergence. By the third iteration, the probabilities become significantly more concentrated as the algorithm increasingly favors near-optimal and optimal solutions, thereby reducing uncertainty and exploratory noise. In the fourth iteration, nearly full convergence is achieved with most of the probability mass focused on the optimal outcome, although slight changes might occur compared to the third iteration. It should be noted that the optimal solution is not always guaranteed to hold the majority of the probability mass, but as the AO iterations progress, the optimal solution consistently emerges as one of the outcomes with the greatest probability mass.

\begin{figure}[t] 
    \centering
    \begin{subfigure}{0.49\columnwidth}
        \centering
        \includegraphics[width=\linewidth]{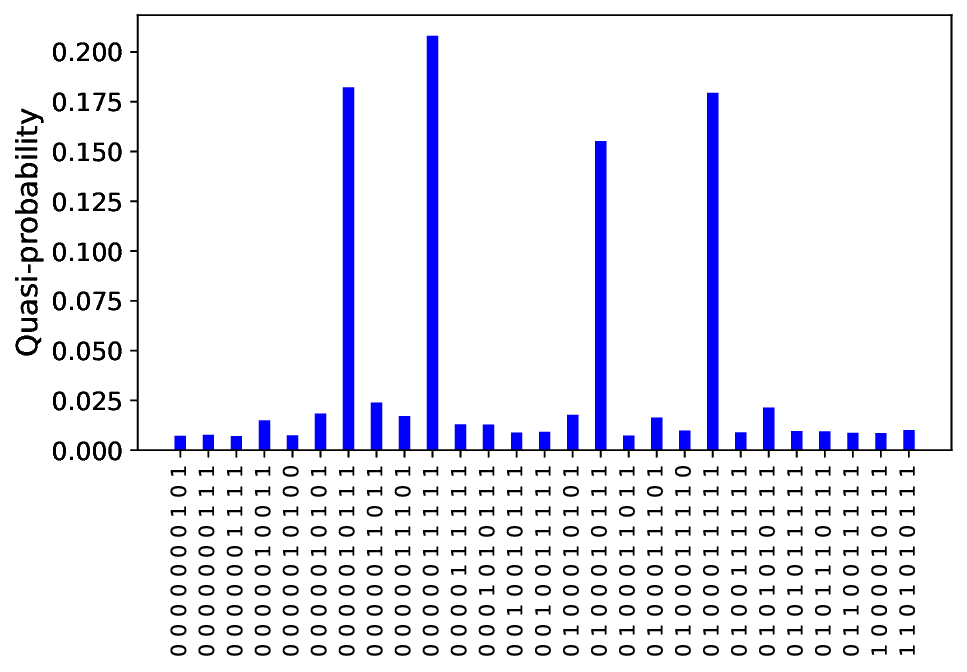}
        \caption{1st AO iteration}
    \end{subfigure} \hfill
    \begin{subfigure}{0.49\columnwidth}
        \centering
        \includegraphics[width=\linewidth]{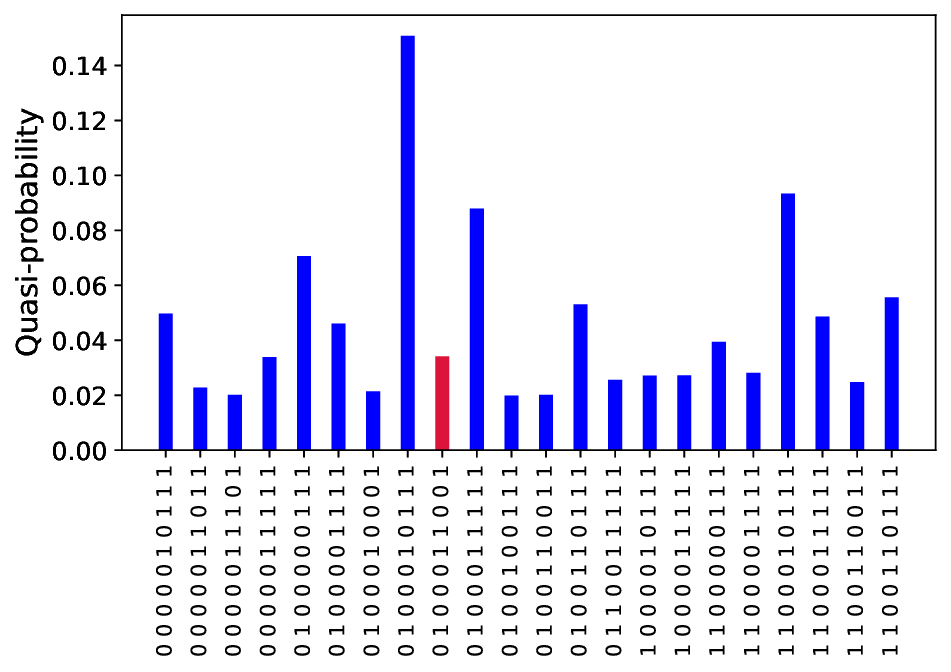}
        \caption{2nd AO iteration}
    \end{subfigure} \\
    \vspace{0.5cm}
    \begin{subfigure}{0.49\columnwidth}
        \centering
        \includegraphics[width=\linewidth]{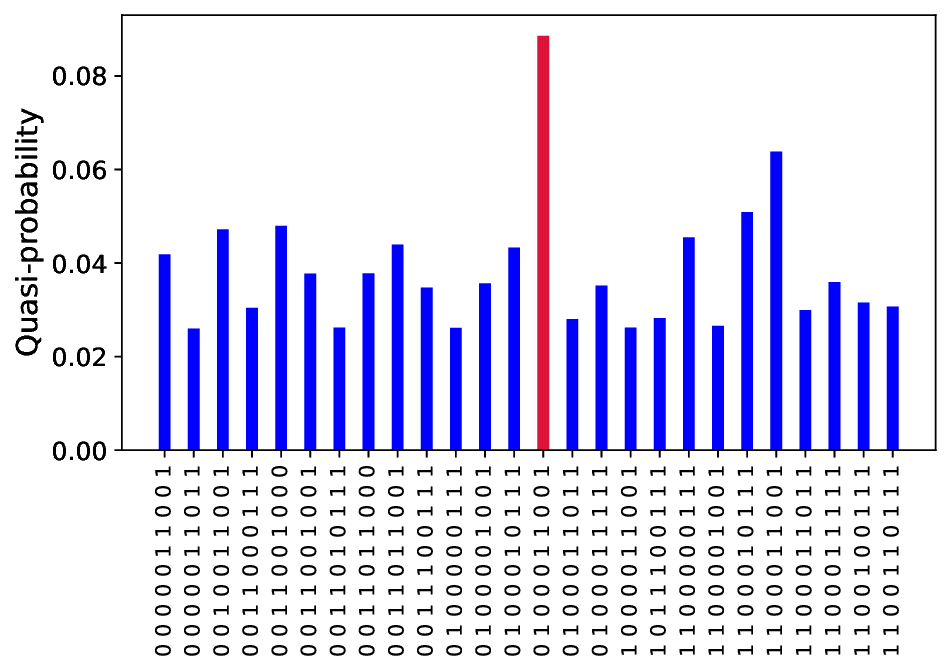}
        \caption{3rd AO iteration}
    \end{subfigure} \hfill
    \begin{subfigure}{0.49\columnwidth}
        \centering
        \includegraphics[width=\linewidth]{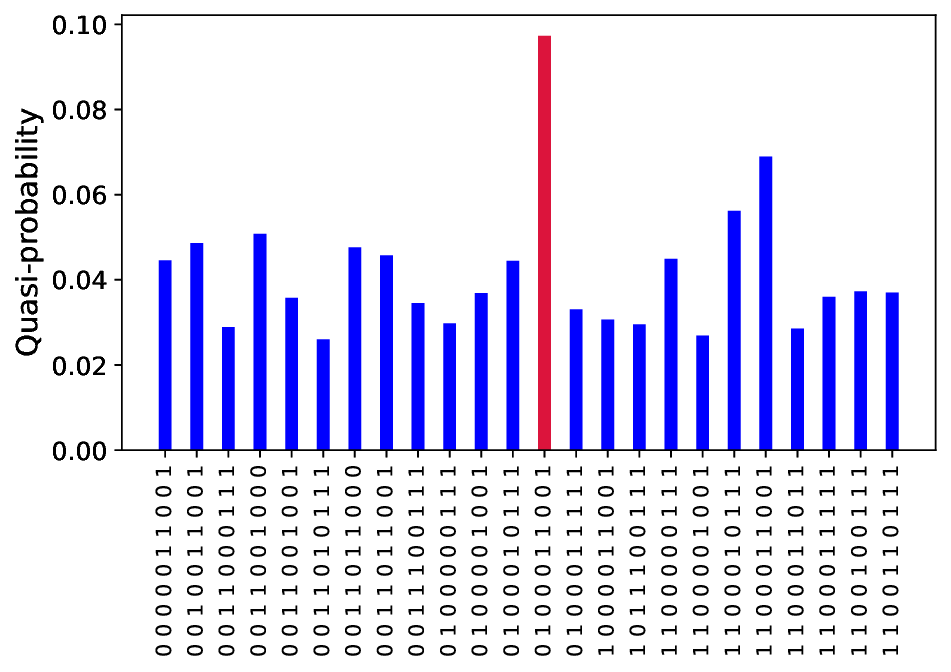}
        \caption{4th AO iteration}
    \end{subfigure}
    \caption{Convergence of the QAOA's quasi-probability distribution.}
    \label{fig:subfigures}
\end{figure}

\section{Conclusions \& Future directions}
In this paper, we addressed the challenge associated with optimizing \(b\)-bit quantized phase shifters in MIMO systems. By leveraging the QAOA and AO, we provided a novel framework for tackling the combinatorial nature of quantized beamforming problems. Our results established a theoretical connection between quantum circuit design and phase shift pre/post-coding, making a significant step toward hybrid-classical optimization in communication systems. Moreover, the utilized warm-start QAOA approach demonstrated enhanced computational efficiency, making it a promising candidate for solving quantized signal processing optimization problems. \textcolor{black}{Numerical experiments confirmed the improved performance of the proposed method against conventional techniques such as quantized SVD and a relaxation-based convex optimization method, showcasing its potential to balance cost-effectiveness and performance. }

Future research includes the exploration of quantum-assisted methods in related fields, such as multi-user MIMO, integrated sensing and communication, and reconfigurable intelligent surfaces. \textcolor{black}{``An interesting approach to reduce the complexity of the proposed method and to confront the qubit-count bottleneck of current hardware, thus scaling the proposed method to a greater number of qubits, is problem decomposition methods, such as circuit knitting. These methods partition the graph corresponding to the original problem into subgraphs that each fit on the available hardware and then combine these partial solutions together to obtain the solution of the initial problem."}
\appendices
\section{Proof of Lemma 1}
A standard approach is to take
\[
  \theta_j 
  \;=\; 
  2^j \,\frac{2\pi}{2^b},
  \quad
  j = 0,1,\dots,b-1.
\]
Note that all these \(\theta_j\) are among the chosen $\phi_i$.
The point is that these \(\theta_j\) will,
through 0/1-sums, produce \emph{all} angles \(\phi_i\). Every integer \(i\), with \( i = 0, ..., 2^b - 1\), has a unique $b$-bit binary expansion as follows
\[
  i
  \;=\;
  x_0\,2^0 \;+\; x_1\,2^1 \;+\;\dots\;+\; x_{b-1}\,2^{b-1},
  \quad 
  x_j\in\{0,1\}.
\]
 Hence, for any \( i = 0, ..., 2^b - 1\) there is a combination of appropriate $x_j\in\{0,1\}$ such that
\[
  \frac{2i\pi}{2^b}
  \;=\; 
  \frac{2\pi}{2^b}
  \Bigl(\sum_{j=0}^{b-1} x_j\,2^j\Bigr).
\]
Thus, by setting \(\theta_j = 2^j\,\frac{\pi}{2^b}\), we have proved that there is always a set of exactly \(b\) angles, \[
  \{\theta_0,\;\theta_1,\;\dots,\;\theta_{b-1}\}\;\subset\;\{\phi_0,\dots,\phi_{2^b-1}\},
\] such that \emph{every} \(\phi_i\) can be written as $
  \phi_i =
  \sum_{j=0}^{b-1} x_j\,\theta_j$.

\bibliographystyle{IEEEtran}
\bibliography{bib}
\end{document}